\documentclass[draft]{agujournal2019}
\usepackage{url} 
\usepackage{lineno}
\usepackage{amsmath, amssymb, amsfonts}
\usepackage[inline]{trackchanges} 
\usepackage{soul}

\usepackage{booktabs}
\usepackage{multirow}

\usepackage{algorithm}
\usepackage{algpseudocode}

\usepackage{placeins}

\draftfalse

\journalname{Journal of Advances in Modeling Earth Systems (JAMES)}

\begin{document}

%
%


\title{Connecting the forward problem to the inverse problem in uncertainty quantification of Earth system models using fast emulators}

%
%




\authors{E. Y. Shin\affil{1}, B. Kale\affil{1}, M. F. Howland\affil{1}}

\affiliation{1}{Civil and Environmental Engineering, Massachusetts Institute of Technology}




\correspondingauthor{ M. F. Howland}{mhowland@mit.edu}



\begin{keypoints}
\item Insights derived from forward uncertainty propagation inform inverse uncertainty quantification of parameterizations in Earth system models
\item We quantify how output sensitivity to parameters depends on quantity, environmental state, time-averaging length, and spatial location
\item We project sensitivity indices onto observation space and use diagnostics to non-iteratively select observations for parameter calibration
\end{keypoints}

%
%

%
%


\begin{abstract}
Quantifying and reducing uncertainty in Earth system model parameterizations is essential to improving their reliability in decision-making. 
Forward uncertainty propagation is used to derive parameter sensitivity but requires physically plausible parameter distributions first be learned from observations. 
Bayesian inference offers a principled approach but can become ill-posed when observations weakly constrain parameters--a condition difficult to know prior to inference. 
Addressing this gap, we show that parameter sensitivity results from forward uncertainty quantification can guide a non-iterative strategy for identifying observations informative to Bayesian calibration. 
We explore both forward and inverse uncertainty quantification for parameterizations of atmospheric turbulence in the Weather Research and Forecasting (WRF) model. 
To overcome the computational bottleneck of $\mathcal{O}(10^5)$ model evaluations required for both analyses, we leverage Gaussian process emulators trained on several hundred WRF simulations. 
Using these emulators, we conduct a global sensitivity analysis across observation space, investigating how parameter contributions to output variance depend on quantity of interest, atmospheric stability, time-averaging length, and spatial location. 
We then introduce nondimensional diagnostic measures that systematically identify regions where a parameter's contribution to output variance exceeds observational noise and its independent effect exceeds interaction effects. 
We demonstrate that observations from these regions serve as a strong proxy for accurate Bayesian calibration and reduced posterior uncertainty. 
Through emulator-aided Bayesian inversion with synthetic observations, we show how parameter uncertainty can be systematically reduced by leveraging sensitivity information. 
Because this framework precedes inference, it provides a practical, non-iterative approach to improving the observational efficiency of Bayesian learning in chaotic dynamical systems.
\end{abstract}

\section*{Plain Language Summary}

Climate and weather models rely on submodels with fixed parameters that can affect their predictions. 
Many of these parameters have traditionally been hand-tuned using limited field data or scientific insight, and their uncertainties are therefore not well known. 
Reducing these uncertainties matters, but it is hard to know ahead of time whether the available data will tell us enough about the parameters we care about. 
Here, we develop a method for choosing data that are most useful for learning uncertain parameters. 
To make this computationally practical, we use artificial intelligence to imitate a submodel for turbulence, allowing us to explore many parameter combinations with only a few hundred model runs instead of millions. 
We then examine how changes in the parameter values affect model predictions. 
This helps reveal how parameter effects differ across atmospheric variables, atmospheric conditions, across space, and across different averaging times. 
Using this information, we identify scenarios where the effects of parameters are strong enough to stand out from observational noise and where the influence of different parameters can be distinguished from one another. 
In a proof-of-concept test using idealized simulations, we observe that parameters with smaller uncertainty learned from data match those expected by our analysis.

%
%

\section{Introduction}
\label{sec:introduction}

Earth system models (ESMs) simulate the Earth's physical processes by solving the governing equations of fluid motion, coupled with biological, chemical, and other physical processes, on a computational grid.
These models make predictions of weather and climate systems, urban flows, pollutant dispersion in complex terrain, and atmosphere–renewable-energy interactions \cite{salamanca2011study, bauer2015quiet, veers2019grand, dunbar2021calibration}.
When processes, which are multi-scale in space and time, fall below the grid scales, submodels, referred to as \emph{parameterizations}, represent the effects of subgrid-scale processes to the grid-scale processes using a combination of empirical and physics-based relationships and empirically calibrated parameters \cite{stensrud2007parameterization}.
A major source of error and uncertainty in parameterizations are these calibrated parameters \cite{stensrud2007parameterization, hourdin2017art, lamb2026perspectives}.
The parameters are often tuned to sparse field observations or idealized simulation data, so they are widely acknowledged to be state-dependent and non-universal.
Quantifying and reducing uncertainty in these parameters is essential to improve predictive accuracy and reliability of ESMs for various applications.

Efforts to quantify uncertainty in empirical parameters have followed two statistical paradigms: the forward approach and the inverse (Bayesian) approach.
The forward approach propagates parameter values through the model to quantify uncertainty in resulting model outputs, which can then be apportioned to the different parameters.
Therefore, in the forward approach, parameter uncertainty is treated as an input assumption, and its influence is quantified through sensitivity analysis.
This has been the classical approach used in previous studies to study uncertainty in ESM parameterizations.
For example, \citeA{nielsen2010evaluation} examined parameter observability and its dependence on observation characteristics, \citeA{yang2017sensitivity} linked parametric sensitivity in planetary boundary layer and surface-layer schemes to terrain and atmospheric stability, and \citeA{baki2022determining} and \citeA{pathak2020uncertainty} applied global sensitivity analysis to identify key parameters across various parameterization schemes.
These studies demonstrate how forward UQ can reveal parameter sensitivity and calibration potential, but they do not determine physically plausible parameter intervals, which must instead be inferred from observations.
The Bayesian inverse approach to UQ provides a principled framework for accounting for observations in a probabilistic manner through posterior distributions over parameters \cite{kennedy2001bayesian}.
This UQ approach has become increasingly popular with the growing availability of data and computational power, and has been applied across multiple scales of Earth system modeling, from climate models \cite{dunbar2021calibration, howland2022parameter} to turbulence \cite{edeling2014bayesian, souza2020uncertainty, shin2026accelerated} and cloud microphysics parameterizations \cite{kim2025uncertainty}.
Yet with a purely Bayesian approach, calibration can become ill-posed if parameters are not well-constrained by the observations or structural model-observation discrepancy dominates \cite{hourdin2017art}.
This issue is typically diagnosed \textit{a posteriori} to the inference process and is directly tied to the observations used for inference.
Therefore, observations play a critical role in the characterization of parameter uncertainty.

Observation selection has been addressed using frameworks such as Bayesian optimal experimental design, in which observations are chosen to maximize the expected information gain about uncertain parameters \cite{chaloner1995bayesian}.
These approaches are formulated using information-theoretic design criteria, such as expected information gain, mutual information or entropy reduction \cite{dunbar2022ensemble, huan2024optimal}.
However, methods based on this framework often rely on iterative and computationally intensive optimization procedures and, in practice, have seen limited adoption due to computational bottlenecks, a lack of standardization, and a focus on methodological development rather than application-specific needs \cite{rainforth2024modern}.
The application of optimal experimental design to the chaotic dynamical atmospheric system is challenging because the informativeness of observations depends on the quantity of interest and on the multi-scale, spatiotemporal variability of the high-dimensional chaotic atmospheric state.
Furthermore, the multi-objective nature of atmospheric modeling and the exhaustive combinatorial search over possible observation subsets makes the resulting optimization computationally infeasible.
As a result, the questions of which variables to observe, how they should be averaged, and where and when observations should be collected to calibrate ESM parameterizations remain critical but largely unanswered.
More generally, the efficient selection of data to improve models is core to model learning across multiple scientific communities \cite{reeves1999sequential,huan2024optimal,lamb2026perspectives}.

In this study, we propose a framework in which global sensitivity analysis based on forward uncertainty propagation guides a non-iterative selection of observations to be used in the Bayesian inverse problem of calibrating and quantifying uncertainty in parameters in a chaotic dynamical system.
To demonstrate the framework, we apply it to parameterizations of turbulence in an atmospheric model.
With the aid of machine-learning emulators, both the forward and inverse uncertainty quantification methods are conducted using only $\mathcal{O}(10^2)$ evaluations of the atmospheric model, in contrast to $\mathcal{O}(10^5)$ model evaluations required in standard sampling-based statistical analysis approaches \cite{geyer2011introduction}.
First, by computing sensitivity indices over the observation space, we show that global sensitivity analysis reveals how parameter sensitivity to key quantities vary across atmospheric stability regimes, time-averaging windows, and spatial observation locations.
Second, we introduce nondimensional diagnostic measures that leverage sensitivity indices to identify regions in the observation space where the parameter signal on model output is stronger than observational noise and the parameter's independent effect is stronger than its interaction effects.
Criteria based on the diagnostic measures enables us to quantify which quantities should be measured, where they should be measured, and how they should be time-averaged in order to construct informative observations of the chaotic dynamical system.
By performing Markov-chain Monte Carlo sampling with the emulators, we show that the diagnostic measures serve as a strong proxy for accurate Bayesian parameter calibration and reduced posterior parameter uncertainty.
Systematic observation selection prior to the inference process improves observational efficiency in Bayesian inference, which is a central topic in inverse problems, parameter estimation, and data assimilation.
Because the statistical methods introduced in this study are not specific to ESM parameterizations, they apply to UQ of models in chaotic dynamical systems more broadly, and are therefore expected to extend beyond the present context to other disciplines.

\section{Methodology}
\label{sec:methodology}

This section outlines the methods used in this study.
First, we detail (1) the global sensitivity analysis to quantify parameter contributions to model-output variance based on forward uncertainty propagation, (2) the Bayesian inverse approach to quantify parameter uncertainty based on observations, (3) the machine-learning emulator that enables both statistical analyses at a feasible computational cost, and (4) a framework that leverages results of the global sensitivity analysis to guide the selection of informative observations for Bayesian inversion.
Then, we describe the atmospheric model, the target parameterizations, and the simulation cases used to generate model outputs and synthetic observations for testing the proposed observation selection framework.

\subsection{Uncertainty quantification}
\label{sec:uncertaintyquantification}

\subsubsection{Global sensitivity analysis}
\label{sec:variancebasedsensitivityanalysismethod}

While sensitivity analysis is formally distinct from an uncertainty analysis, which characterizes uncertainty in model prediction \cite{saltelli2019so}, the present work focuses on quantifying uncertainty in model parameters. 
In this context, sensitivity analysis is a method within the forward uncertainty quantification paradigm, as it forward-propagates values from a prescribed parameter space to determine how uncertainty in the outputs can be attributed to the parameters.
Global sensitivity analysis quantifies how the variance in the model output can be apportioned to the input parameters, in the form of main, interaction, and total effects \cite{saltelli2002sensitivity}.

Consider a general model $\mathcal{G}(\boldsymbol{X})$, where 
$\boldsymbol{X} = (X_1, \ldots, X_p)$ denotes a set of $p$ independent parameters. 
Let the observable model output be defined as
\[
f(\boldsymbol{X}) = \mathcal{H}(\mathcal{G}(\boldsymbol{X})),
\]
where $\mathcal{H}$ denotes the observation operator that maps the model output to the observation space. 
The $f(\boldsymbol{X})$ is assumed to be square-integrable with respect to the joint probability distribution of $\boldsymbol{X}$, i.e., $\mathbb{E}[f(\boldsymbol{X})^2] < \infty$.
Under these assumptions, the analysis of variance (ANOVA) decomposition allows the variance of the model output to be written as
\begin{align}
\operatorname{Var}[f(\boldsymbol{X})] 
&= \sum_{i=1}^{p} \operatorname{Var}[f_i(X_i)]
+ \sum_{i < j}^{p} \operatorname{Var}[f_{ij}(X_i,X_j)]
+ \cdots 
+ \operatorname{Var}[f_{1,2,\ldots,p}(X_1,X_2,\ldots,X_p)],
\end{align}
where the components are orthogonal contributions from parameters and their interactions.
Based on this variance decomposition, three types of sensitivity indices can be constructed--each representing a different contribution of a parameter $X_i$ to the variance of the model output $f(\boldsymbol{X})$.

The first-order, or main-effect, sensitivity of parameter $X_i$, which quantifies its independent contribution to model-output variance excluding interactions with other parameters, is defined as
\begin{equation}
S_i =
\frac{
\operatorname{Var}_{X_i}
\left[
\mathbb{E}_{\boldsymbol{X}_{\sim i}}
\left[
f(\boldsymbol{X}) \mid X_i
\right]
\right]
}{
\operatorname{Var}\!\left[f(\boldsymbol{X})\right]
}.
\end{equation}

The second-order, or interaction-effect, sensitivity of parameter $X_i$, which quantifies the contribution solely from the pairwise interaction between $X_i$ and $X_j$, is defined as
\begin{equation}
S_{ij} =
\frac{
\operatorname{Var}_{X_{ij}}
\!\left[
\mathbb{E}_{\boldsymbol{X}_{\sim ij}}
\!\left[
f(\boldsymbol{X}) \mid X_{ij}
\right]
\right]
-
\operatorname{Var}_{X_i}
\!\left[
\mathbb{E}_{\boldsymbol{X}_{\sim i}}
\!\left[
f(\boldsymbol{X}) \mid X_i
\right]
\right]
-
\operatorname{Var}_{X_j}
\!\left[
\mathbb{E}_{\boldsymbol{X}_{\sim j}}
\!\left[
f(\boldsymbol{X}) \mid X_j
\right]
\right]
}{
\operatorname{Var}\!\left[f(\boldsymbol{X})\right]
}.
\end{equation}

The total-effect sensitivity of parameter $X_i$, which quantifies its contribution to output variance including all interactions with other parameters, is defined as
\begin{equation}
S_{T_i} =
\frac{
\mathbb{E}_{\boldsymbol{X}_{\sim i}}
\!\left[
\operatorname{Var}_{X_i}
\!\left[
f(\boldsymbol{X}) \mid \boldsymbol{X}_{\sim i}
\right]
\right]
}{
\operatorname{Var}\!\left[f(\boldsymbol{X})\right]
}
=
1 -
\frac{
\operatorname{Var}_{\boldsymbol{X}_{\sim i}}
\!\left[
\mathbb{E}_{X_i}
\!\left[
f(\boldsymbol{X}) \mid \boldsymbol{X}_{\sim i}
\right]
\right]
}{
\operatorname{Var}\!\left[f(\boldsymbol{X})\right]
}.
\end{equation}

These three contributions to sensitivity of parameter $X_i$ are related to each other and higher-order terms as 
\begin{equation}
S_{T_i}
=
S_i
+ \sum_{j \ne i} S_{ij}
+ \sum_{\substack{j<k\\ j,k \ne i}} S_{ijk}
+ \cdots
+ S_{1,\ldots,p},
\label{eq:ST_expand}
\end{equation}
where the sensitivity indices can be interpreted as fractions of the total model-output variance $\operatorname{Var}\!\left[f(\boldsymbol{X})\right]$.
For example, the main-effect index $S_i$ represents the fraction of the total model-output variance attributable to the independent contribution of parameter $X_i$, whereas the total-effect index $S_{T_i}$ represents the fraction attributable to all contributions involving that parameter, including both its independent effect and its interactions with other parameters.

These formulations of the main-, interaction-, and total-effect Sobol’ sensitivity indices are computed following \citeA{saltelli2002making} and \citeA{saltelli2010variance}.
We use \textit{SALib}, an open-source Python library for sensitivity analysis \cite{Herman2017,Iwanaga2022}.
Two independent sample matrices $\mathbf A,\mathbf B\in\mathbb R^{N\times p}$ are drawn, where $N$ is the number of parameter-set combinations or samples.
We define $f(\mathbf A),f(\mathbf B)\in\mathbb R^{N}$ to be the observable model outputs, obtained by propagating each row of $\mathbf A$ and $\mathbf B$ through the model.
For each parameter $i$, we form the hybrid matrix $\mathbf A_{\mathbf B}^{(i)}$ by replacing the $i$th column of $\mathbf A$ with the corresponding column of $\mathbf B$, and evaluate $f(\mathbf A_{\mathbf B}^{(i)})$.
For second-order interaction effects between parameters $i$ and $j$, we use the hybrid matrices $\mathbf A_{\mathbf B}^{(j)}$ and $\mathbf B_{\mathbf A}^{(i)}$, obtained by replacing the $j$th column of $\mathbf A$ with that of $\mathbf B$, and the $i$th column of $\mathbf B$ with that of $\mathbf A$, respectively.
Computing the main- and total-effect indices requires at minimum $N(p + 2)$ model evaluations, whereas additionally computing interaction-effect indices increases the cost to at minimum $N(2p + 2)$ model evaluations.
The Saltelli estimators for the variance components are summarized as
\begin{align}
&
\operatorname{Var}_{X_i}
\!\left[
\mathbb{E}_{\boldsymbol{X}_{\sim i}}
\!\left[
f(\boldsymbol{X}) \mid X_i
\right]
\right]
\approx
\frac{1}{N}
\sum_{n=1}^{N}
f(\mathbf B)_n
\left[
f\!\left(\mathbf A_\mathbf B^{(i)}\right)_n
-
f(\mathbf A)_n
\right],
\\
&
\operatorname{Var}_{X_{ij}}
\!\left[
\mathbb{E}_{\boldsymbol{X}_{\sim ij}}
\!\left[
f(\boldsymbol{X}) \mid X_{ij}
\right]
\right]
\approx
\frac{1}{N}
\sum_{n=1}^{N}
\left[
f\!\left(\mathbf B_\mathbf A^{(i)}\right)_n
f\!\left(\mathbf A_\mathbf B^{(j)}\right)_n
-
f(\mathbf{A})_n
f(\mathbf{B})_n
\right]
\\
&
\mathbb{E}_{\boldsymbol{X}_{\sim i}}
\!\left[
\operatorname{Var}_{X_i}
\!\left[
f(\boldsymbol{X}) \mid \boldsymbol{X}_{\sim i}
\right]
\right]
\approx
\frac{1}{2N}
\sum_{n=1}^{N}
\left[
f(\mathbf A)_n
-
f\!\left(\mathbf A_\mathbf B^{(i)}\right)_n
\right]^2.
\end{align}
where $n=1,\dots,N$ indexes the samples.
These estimators can be computed independently for any model output, with an accuracy that depends on sufficient sampling (high $N$) of the parameter space.
The primary bottleneck of global sensitivity analysis is the large number ($\sim \mathcal{O}(10^5)$) of model evaluations required to converge the numerically computed sensitivity indices, especially for highly non-linear systems \cite{saxton2024convergence}.

\subsubsection{Bayesian inversion}
\label{sec:bayesianinversion}

The probabilistic learning of model parameters from observations may be formulated as a Bayesian inverse problem.
Here, we derive the inverse formulation useful for parameter estimation in a chaotic dynamical system.
The relationship between the model and observation is given as
\begin{equation}
\boldsymbol{y}(\boldsymbol{\xi}^{(i)})
=
\mathcal{H}(\mathcal{G}(\boldsymbol{X},\boldsymbol{\xi}^{(j)}))
+
\boldsymbol{\delta}
+
\boldsymbol{\eta},
\end{equation}
where $\boldsymbol{y}$ is the observation, $\mathcal{G}$ the model, $\mathcal{H}$ is the observation operator, $\boldsymbol{X}$ the parameters, $\boldsymbol{\xi}$ the initial conditions in $(i)$ observation and $(j)$ model that do not necessarily match, $\boldsymbol{\eta}$ the measurement error, and $\boldsymbol{\delta}$ the model-form error.
The inverse mapping is ill-posed for the identification of parameters since it necessitates both state and parameter estimation in the chaotic system.
For well-posed parameter estimation, we first assume no model--data discrepancy ($\boldsymbol{\delta}=0$) and negligible measurement error ($\boldsymbol{\eta}=0$), as this study uses synthetic observations generated by the target model.

Let $\boldsymbol{y}_\infty$ denote the infinite-time (or ensemble) average of the observation. 
Under ergodic assumptions, sufficiently long time averaging removes the influence of transient initial conditions in both the model and observation, yielding the relationship
\begin{equation}
\boldsymbol{y}_\infty = \mathcal{H}( \mathcal{G}_\infty(\boldsymbol{X}) ),
\label{infinite}
\end{equation}
where $\mathcal{G}_\infty$ denotes the corresponding infinite-time average of the model output. 
This provides a well-posed inverse problem for the parameters $\boldsymbol{X}$.
In practice, however, infinite averages are not available. Instead, both model output and observation are computed as finite-time averages.
We assume that the finite time-averaged model and observations approximate their corresponding infinite-time averages, with the effects of internal variability in each represented as Gaussian noise.
Using the central limit theorem, the finite-time averaged observation $\overline{\boldsymbol{y}}$ can be approximated as
\begin{equation}
\overline{\boldsymbol{y}} (\boldsymbol{\xi}^{(i)}) \approx \boldsymbol{y}_\infty + \gamma,
\qquad
\gamma \sim \mathcal{N}(0,\Gamma),
\label{finite_y}
\end{equation}
where the overbar denotes time-averaging, and $\Gamma$ is the observational noise covariance matrix arising from finite averaging of the observations.
Similarly, the finite-time average of the model is given as
\begin{equation}
\overline{\mathcal{G}}(\boldsymbol{X},\boldsymbol{\xi}^{(j)})
\approx
\mathcal{G}_\infty(\boldsymbol{X}) + \sigma,
\qquad
\sigma \sim \mathcal{N}(0,\Sigma),
\label{finite_G}
\end{equation}
where $\Sigma$ represents the noise covariance associated with finite-time averaging of the model.
In other words, realizations of $\overline{\mathcal{G}}(\boldsymbol{X},\boldsymbol{\xi}^{(j)})$ and $\overline{\boldsymbol{y}} (\boldsymbol{\xi}^{(i)})$ both still contain internal variability of the chaotic system. 

Combining Eqs.~(\ref{infinite}) and (\ref{finite_y}) yields an inverse formulation using the finite-time averaged observation and infinite-time averaged model
\begin{equation}
    \overline{\boldsymbol{y}} (\boldsymbol{\xi}^{(i)})=\mathcal{H}(\mathcal{G}_\infty(\boldsymbol{X}))+\gamma,
\end{equation}
where initial conditions in the model and observation no longer have to match, and error $\gamma$ is due to finite averaging of the observation.
For brevity, we omit the overbar on the observation and adopt the notation 
$f_\infty (\boldsymbol{X}) = \mathcal{H}(\mathcal{G}_\infty(\boldsymbol{X}))$ as in the global sensitivity analysis, to rewrite the final formulation of the inverse problem as
\begin{equation}
\boldsymbol{y} = f_\infty (\boldsymbol{X}) + \gamma,
\label{eq:inverseproblem}
\end{equation}
which solves for the infinite-time-averaged model and finite-time averaged observation.
We note that while the infinite-time averaged model $f_\infty(\boldsymbol{X})$ cannot be evaluated directly, the emulator detailed in Sect.~\ref{sec:machinelearningemulation} provides a surrogate approximation because independent finite-averaged model realizations $\mathcal{H}(\overline{\mathcal{G}}(\boldsymbol{X},\boldsymbol{\xi}^{(j)}))$ used as emulator training data sample different instantiations of internal variability $\mathcal{H} \Sigma\mathcal{H}^{\top}$ in the modeled chaotic system.

In the inverse problem (Eq.~\ref{eq:inverseproblem}), learning parameters $\boldsymbol{X}$ from observations $\boldsymbol{y}$ relies on Bayes' Theorem
\begin{equation}
  p(\boldsymbol{X}\mid \boldsymbol{y})
  =
  \frac{p(\boldsymbol{y}\mid \boldsymbol{X})\,p(\boldsymbol{X})}
       {\displaystyle \int p(\boldsymbol{y}\mid \boldsymbol{X})\,p(\boldsymbol{X})\,\mathrm{d}\boldsymbol{X}},
\end{equation}
where $p(\boldsymbol{X} \mid \boldsymbol{y})$ is the posterior, $p(\boldsymbol{y} \mid \boldsymbol{X})$ the model-observation likelihood, and $p(\boldsymbol{X})$ the prior distributions of the parameters.
Except in a few conjugate cases, the posterior has no closed form, and the marginal likelihood is an intractably high-dimensional integral.
Therefore, the posterior distribution is computed using stochastic sampling methods, such as Markov chain Monte Carlo (MCMC) \cite{robert1999monte}.
In this study, because the goal is not to demonstrate the latest variant of MCMC sampling methods, we use the well-established and widely used Metropolis-Hastings algorithm to sample the posterior \cite{hastings1970monte}.
Up to a normalization constant, the Bayesian posterior distribution is given as
\begin{equation}
p(\boldsymbol{X}\mid\boldsymbol{y})
\propto
\exp\!\left(
-\frac{1}{2}
\left\|
\boldsymbol{y}- f_\infty(\boldsymbol{X})
\right\|_{\Gamma}^{2}
\right) p(\boldsymbol{X}),
\end{equation}
corresponding to a Gaussian likelihood with covariance $\Gamma$, i.e., the observational noise covariance.
As in global sensitivity analysis, the key bottleneck in Bayesian model inversion is the large number ($\sim \mathcal{O}(10^5)$) of model evaluations to accurately characterize the joint posterior distribution of model parameters \cite{kennedy2001bayesian}.

\subsubsection{Machine-learning emulation}
\label{sec:machinelearningemulation}

Sampling-based statistical analyses are constrained by the computational cost of the large number of model evaluations required for the numerical computation of statistical measures.
As a remedy, machine-learning (ML) emulators act as data-efficient surrogates of the expensive numerical models \cite{sacks1989design,forrester2008engineering}.
While innovative methods continue to improve the predictive accuracy and generalizability of ML emulators, we rely here on the strong interpolative accuracy of Gaussian process emulators, especially since training data are limited ($\sim \mathcal{O}(10^2)$ input-output pairs).
The emulator is trained on pairs of parameter sets and their corresponding model outputs.
Its performance is evaluated using a standard 80–20 train–test split (see \ref{sec:appendix1}).
We sample the computationally cheap emulator in place of the true atmospheric model to enable global sensitivity analysis and Bayesian inversion of parameters.
The emulator is a modular component and can be replaced with other ML tools (e.g., neural networks) as needed in other application settings.

Gaussian processes (GP) are widely used as emulators for sampling and uncertainty propagation in settings where only a modest number of training data are available \cite{hoolohan2018improved,gramacy2020surrogates}.
A GP is a distribution over functions such that any finite collection of function values follows a multivariate Gaussian distribution.
It is fully characterized by a mean function $m(\boldsymbol{X})$ and a covariance function $\kappa(\boldsymbol{X},\boldsymbol{X}')$.
In this work, we construct GP emulators to represent the mapping between uncertain model parameters and quantities of interest depending on the spatial position.
Let $\boldsymbol{X}\in\mathbb{R}^p$ denote the vector of input parameters, and let
$f(\boldsymbol{X}) = [f_1(\boldsymbol{X}),\ldots,f_q(\boldsymbol{X})]^{\top} \in\mathbb{R}^q$ denote the vector of model outputs corresponding to a specific quantity of interest evaluated at $q$ spatial locations $s_1,...,s_q$.
All inputs and outputs are first standardized to zero mean and unit variance using statistics computed from the training set.
Because outputs at adjacent spatial locations may be physically correlated, we apply a Principal Component Analysis (PCA) decomposition to the standardized training outputs.

Given a set of $M$ training simulations, we form the output matrix $\boldsymbol{F}\in\mathbb{R}^{M\times q}$ and compute its PCA decomposition
\begin{equation}
  f(\boldsymbol{X})
    \approx \boldsymbol{\mu}_f
    + \sum_{k=1}^{K} \alpha_k(\boldsymbol{X})\,\boldsymbol{\phi}_k,
\end{equation}
where $\boldsymbol{\mu}_f\in\mathbb{R}^q$ is the sample mean, $\boldsymbol{\phi}_k\in\mathbb{R}^q$ are the principal component basis vectors, and $\alpha_k(\boldsymbol{X})\in\mathbb{R}$ are the corresponding projection coefficients.
We place independent GP priors on each coefficient function,
\begin{equation}
  \alpha_k(\boldsymbol{X})
    \sim \mathcal{GP}\!\left(0,\;
         \kappa_k(\boldsymbol{X},\boldsymbol{X}')\right),
  \quad k = 1,\ldots,K,
\end{equation}
so that for a vector of input parameters $\boldsymbol{X}^*$, each GP yields a predictive mean $\hat{\alpha}_k(\boldsymbol{X}^*)$ and variance $\sigma^2_{\alpha_k}(\boldsymbol{X}^*)$.
The choice of the covariance kernel is application-dependent. 
Here, the covariance structure is specified using a Mat\'ern kernel with an additive white-noise kernel to account for numerical noise.
The full-profile predictive mean and covariance are then reconstructed as
\begin{equation}
    f_{\mathrm{GP}}(\boldsymbol{X}^*)
    = \boldsymbol{\mu}_f
    + \sum_{k=1}^{K}
      \hat{\alpha}_k(\boldsymbol{X}^*)\,\boldsymbol{\phi}_k,
    \qquad
    \Sigma_{\mathrm{GP}}(\boldsymbol{X}^*)
    = \sum_{k=1}^{K}
      \sigma^2_{\alpha_k}(\boldsymbol{X}^*)\,
      \boldsymbol{\phi}_k\,\boldsymbol{\phi}_k^{\top}.
\end{equation}
This allows the constructed emulator to preserve spatially coherent predictions and uncertainty estimates.
In this study, all $K$ modes are retained to preserve the full variability in the training data.

Although the infinite-time averaged model $f_\infty(\boldsymbol{X})$ is not directly available, the emulator, once trained on independent samples of the finite-time-averaged model outputs that sample different instantiations of internal variability, approximates a predictive distribution for the infinite-average model output \cite{dunbar2021calibration, howland2022parameter}
\begin{equation}
    f_\infty(\boldsymbol{X}) \approx 
    \mathcal{N} (f_{\mathrm{GP}}(\boldsymbol{X}),   \Sigma_{\mathrm{GP}}(\boldsymbol{X})).
\end{equation}

Using the emulator as a surrogate for the forward model in global sensitivity analysis, we compute Saltelli estimators of the Sobol variance components as
\begin{align}
&
\operatorname{Var}_{X_i}
\!\left[
\mathbb{E}_{\boldsymbol{X}_{\sim i}}
\!\left[
f_{\mathrm{GP}}(\boldsymbol{X}) \mid X_i
\right]
\right]
\approx
\frac{1}{N}
\sum_{n=1}^{N}
f_{\mathrm{GP}}(\mathbf B)_n
\left[
f_{\mathrm{GP}}(\mathbf A_\mathbf B^{(i)})_n
-
f_{\mathrm{GP}}(\mathbf A)_n
\right],
\\
&
\operatorname{Var}_{X_{ij}}
\!\left[
\mathbb{E}_{\boldsymbol{X}_{\sim ij}}
\!\left[
f_{\mathrm{GP}}(\boldsymbol{X}) \mid X_{ij}
\right]
\right]
\approx
\frac{1}{N}
\sum_{n=1}^{N}
\left[
f_{\mathrm{GP}}\!\left(\mathbf B_\mathbf A^{(i)}\right)_n
f_{\mathrm{GP}}\!\left(\mathbf A_\mathbf B^{(j)}\right)_n
-
f_{\mathrm{GP}}(\mathbf{A})_n
f_{\mathrm{GP}}(\mathbf{B})_n
\right]
\\
&
\mathbb{E}_{\boldsymbol{X}_{\sim i}}
\!\left[
\operatorname{Var}_{X_i}
\!\left[
f_{\mathrm{GP}}(\boldsymbol{X}) \mid \boldsymbol{X}_{\sim i}
\right]
\right]
\approx
\frac{1}{2N}
\sum_{n=1}^{N}
\left[
f_{\mathrm{GP}}(\mathbf A)_n
-
f_{\mathrm{GP}}(\mathbf A_\mathbf B^{(i)})_n
\right]^2.
\end{align}
where $\mathbf A$ and $\mathbf B$ are two independent sample matrices drawn
from the input distribution, and $\mathbf A_\mathbf B^{(i)}$ and $\mathbf B_\mathbf A^{(j)}$ are the hybrid matrices obtained by replacing the $i$-th column of $\mathbf A$ with the corresponding column of $\mathbf B$, and the $j$-th column of $\mathbf B$ with the corresponding column of $\mathbf A$, respectively.

Using the emulator in Bayesian inference, we approximate the posterior distribution as
\begin{equation}
p(\boldsymbol{X} \mid\boldsymbol{y})
\propto
\exp \left(
-\frac{1}{2}
\left\|
\boldsymbol{y}-f_{\mathrm{GP}}(\boldsymbol{X})
\right\|^2_{\Gamma+\Sigma_{\mathrm{GP}}(\boldsymbol X)}
-\frac{1}{2}
\log\det(\Gamma+\Sigma_{\mathrm{GP}}(\boldsymbol X))
\right) p(\boldsymbol{X}).
\end{equation}
which corresponds to a Gaussian likelihood whose covariance is the sum of the observational covariance and the GP predictive covariance.

\subsection{Observation selection framework}
\label{sec:observationselectionframework}

In the chaotic atmospheric system, the informativeness of observations varies across space and time within a high-dimensional state space.
While Bayesian optimal experimental design provides a principled way to account for observational uncertainty and select informative measurements, the inherently multi-objective nature of atmospheric modeling and the combinatorial search over possible measurement subsets makes the resulting optimization often computationally infeasible in practice.
To avoid an iterative optimization process, we propose a framework to systematically and efficiently identify informative observations for the Bayesian inverse problem of parameter calibration and uncertainty reduction.
We note that the informativeness of an observation in model calibration is its value \textit{to the model}, and in this application, to the model parameters.
Therefore, we assert that parameter sensitivity projected onto the observation space is the primary indicator for the informativeness an observation for parameter calibration.
This motivates our use of nondimensional diagnostic measures of Sobol' sensitivity indices that are computed over the observation space.

\begin{figure}
    \centering
        \includegraphics[width=1.0\textwidth]{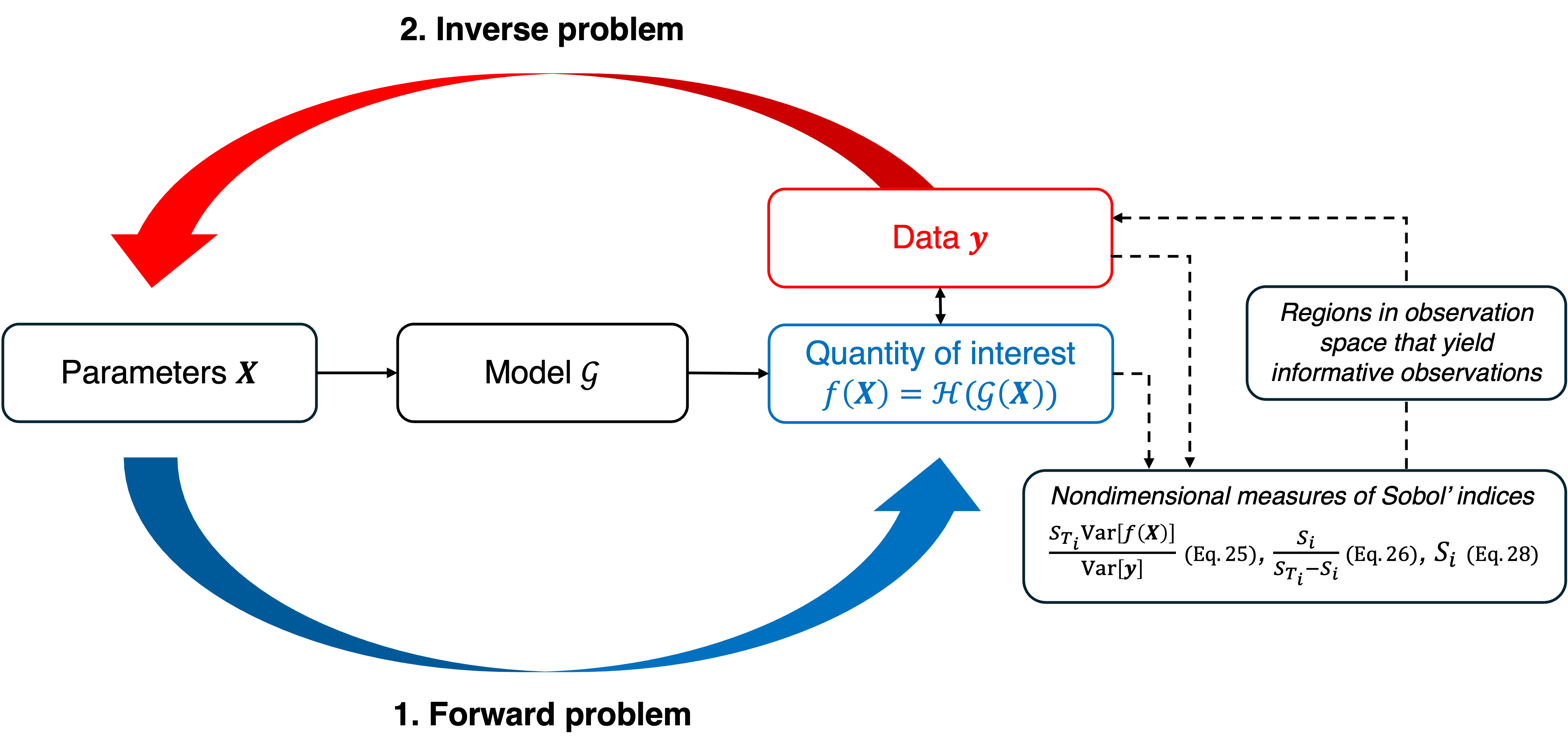}
    \caption{
    Schematic of the proposed framework that connects the forward problem to the inverse problem of UQ through nondimensional measures of Sobol’ sensitivity indices.
    These diagnostics quantify (i) the total parameter signal relative to observational noise, (ii) the ratio of main-effect sensitivity of a parameter to its interaction-effect sensitivity, and (iii) the main-effect sensitivity, to identify regions in the observation space that yield informative observations.
    }
    \label{fig:framework_schematic}
\end{figure}

The diagnostic measures, which are derived from the forward problem and aim to guide the inverse problem of UQ, are illustrated in Fig.~\ref{fig:framework_schematic}.
Within the inverse problem defined by Eq.~\ref{eq:inverseproblem}, we focus on two key challenges to parameter calibration in general, which are also especially important to ESM applications: (i) observational noise and (ii) confounding effects from other parameters.
Parameter calibration is only possible when the parameter’s signal in the model output is distinguishable from the observation noise.
Based on the decomposition of variance of the model output in global sensitivity analysis, we define a parameter's total signal as the dimensional total-effect contribution of $X_i$ to the model-output variance, given as $S_{T_i}\operatorname{Var}[f(\boldsymbol X)]$.
The corresponding observational noise, or variance in observation, is expressed as $\operatorname{Var}[\boldsymbol{y}]$.
We define the comparison of the two quantities as the signal-to-noise (SNR) ratio
\begin{align}
\operatorname{SNR}_\mathcal{A}(i)
=
\frac{S_{T_i}\,\operatorname{Var}[f(\boldsymbol X)]}
{\operatorname{Var}[\boldsymbol{y}]},
\label{eq:snr}
\end{align}
where $i$ indexes the parameter of interest, and $\mathcal{A}$ represents the observation space defined by the observation operator $\mathcal{H}$.
We use the criterion
\begin{align*}
    \operatorname{SNR}_\mathcal{A}(i) \ge 1 
    &\;\Rightarrow\; \text{$X_i$'s signal larger than observational noise}, \\
    \operatorname{SNR}_\mathcal{A}(i) < 1 
    &\;\Rightarrow\; \text{$X_i$'s signal  smaller than observational noise}.
\end{align*}
Thus, $\operatorname{SNR}_\mathcal{A}(i) \ge 1$ is a necessary, but not sufficient, condition for calibrating $X_i$ using observations in observation space $\mathcal{A}$.

Interactions among parameters can produce compensating or redundant effects through which multiple parameter values can yield similar model outputs \cite{lam2022practical}.
This can make it difficult to uniquely identify parameter values during calibration.
Therefore, we aim to identify regions in the observation space where independent parameter effects are stronger than corresponding parameter interaction effects.
From parameters whose sensitivities are distinguishable from observational noise based on the signal-to-noise ratio, we choose to compare the main-effect sensitivity $S_i$ against the full interaction-effect sensitivity $S_{T_i} - S_{i}$.
We define this comparison as the main-to-interaction (MIR) ratio
\begin{align}
  \operatorname{MIR}_{\mathcal{A}} (i)
  =
  \frac{S_{i}}
       { S_{T_i} - S_{i} },
\label{eq:mir}
\end{align}
which is effectively a comparison of sensitivity contributions \textit{local} to parameter $X_i$.
We use the criterion
\begin{align*}
    \operatorname{MIR}_{\mathcal{A}} (i) \ge 1 
    &\;\Rightarrow\; \text{$X_i$'s independent effect larger than interaction effects}, \\
    \operatorname{MIR}_{\mathcal{A}} (i) < 1 
    &\;\Rightarrow\; \text{$X_i$'s interaction effects larger than independent effect}.
\end{align*}
The condition $\operatorname{MIR}_{\mathcal{A}} (i) \ge 1$ aims to identify regions in the observation space where the parameter sensitivity is primarily driven by independent effects rather than interaction effects.

The two nondimensional diagnostic measures (Eqs.~\ref{eq:snr} and \ref{eq:mir}) identify regions in the observation space where the parameter signal in model output exceeds observational noise and is minimally influenced by interactions with other parameters.
This provides an alternative method for diagnosing the practical identifiability of parameters \cite{wieland2021structural, lam2022practical}.
In contrast to the established profile-likelihood method for diagnosing practical identifiability \cite{raue2009structural}, our approach projects parameter sensitivity onto the observation space, which provides a visualization of an informative observation space prior to any optimization process.
We note practical identifiability may be interpreted as a condition of \textit{sufficiency} to constrain parameters.
This represents a lower bound on the informativeness of observations with respect to parameter $X_i$, given as
\begin{align}
\mathcal{A}(i)^{\mathrm{lower}}
=
\left\{
\mathcal{A}
:\;
\operatorname{SNR}_{\mathcal{A}} (i) \ge 1,\;
\operatorname{MIR}_{\mathcal{A}} (i) \ge 1\
\right\}.
\label{eq:lower}
\end{align}

An upper bound on informativeness could, in principle, be defined through an optimality criterion; however, formal notions of optimality are difficult to practically utilize in a spatiotemporally varying high-dimensional state space.
Instead, we adopt a sensitivity-based definition of an upper bound on informativeness.
The independent contribution of a parameter relative to all contributions by parameters is a \textit{global} comparison in the parameter space and is simply the main-effect sensitivity
\begin{align}
\frac{S_{i}}
{
\sum_{i} S_i + \sum_{i<j} S_{ij} + \cdots + S_{1,2,\ldots,p}
}
= S_{i}.
\label{eq:main_to_total}
\end{align}

Since the parameter must first be identifiable over observational noise, the upper bound on informativeness for parameter $X_i$ is attained at the region in the observation space where the following condition is satisfied 
\begin{align}
\mathcal{A}(i)^{\mathrm{upper}}
=
\left\{
\mathcal{A}
:\;
\operatorname{SNR}_{\mathcal{A}} (i) \ge 1,\;
\max_{\mathcal{A}} S_{i,\mathcal{A}}
\right\}.
\label{eq:upper}
\end{align}

To summarize, the three nondimensional diagnostic measures based on Sobol' sensitivity indices identify regions in the observation space where the total parameter signal exceeds observational noise (Eq.~\ref{eq:snr}), the main-effect sensitivity exceeds the full interaction-effect sensitivity local to the parameter (Eq.~\ref{eq:mir}), and the main-effect sensitivity is globally maximum in the parameter space (Eq.~\ref{eq:main_to_total}).

\subsubsection{Limitations of the framework}
The nondimensional diagnostic measures inherit assumptions from the ANOVA decomposition used in global sensitivity analysis.
Their reliability depends on the accuracy of the statistical estimators, which in this study is determined by (i) the number of emulator samples used to compute the estimators and (ii) the number of model evaluations used to train the emulator.
Furthermore, statistical analysis of the parameter-model-output mapping is of an interpolative nature and is not guaranteed to generalize outside the sampled parameter space.
Lastly, in practice, the overlap region identified by the diagnostic measures may be empty or may result in multiple disjoint regions, reflecting competing trade-offs in calibrating multiple parameters.

\subsection{Atmospheric model}
\label{sec:forwardmodel}

Here, we introduce the atmospheric model we use to perform the UQ methods and test the observation selection framework.
In this study, we use the Weather Research and Forecasting (WRF) Model v4.6.1 \cite{skamarock2019description}.
WRF consists of a nonhydrostatic, compressible dynamical core with configurable, coupled physics parameterizations, which allow the representation of atmospheric processes across multiple spatiotemporal scales.
We focus on the parameterization of turbulent processes in the planetary boundary layer (PBL)--the layer of the troposphere that is directly influenced by the presence of the surface and its forcings \cite{stull2012introduction}.
Terminology for the parameterization of turbulence varies across scientific communities--\emph{turbulence model}, \emph{turbulence parameterization}, and \emph{PBL scheme} are a few terms that are used interchangeably.
Here, we target the PBL scheme and, additionally, the surface-layer (SFCLAY) scheme, which parameterizes the surface momentum and heat fluxes used by the PBL scheme.
To eliminate confounding of time-varying mesoscale forcings and surface heterogeneity from vertical turbulent processes by the PBL scheme, we use idealized, horizontally homogeneous, dry PBL simulations throughout this study.

\subsubsection{Planetary boundary layer and surface layer schemes}
\label{sec:pblsfclayschemes}

We focus on the Yonsei University (YSU) PBL scheme \cite{hong2006new}, coupled with the revised version of the fifth-generation Pennsylvania State University-National Center for Atmospheric Research Mesoscale Model (revised MM5) SFCLAY scheme \cite{jimenez2012revised}.
This combination is widely used and therefore practically relevant for UQ.
Results for two other PBL schemes are provided in \ref{sec:appendix2}, demonstrating similar qualitative results, and are therefore not emphasized in the main text for brevity.
The YSU PBL scheme is a first-order nonlocal K-theory turbulence closure with an added upward mass-flux term to represent countergradient transport.
We target the parameter set $\boldsymbol{X} = \{a, b, p, d_1, d_2, z_0\}$, which are listed in Table~\ref{tab:sl-pbl-parameters}.
The two right-hand columns show the standard values of the parameters used in the model and the initial parameter ranges prescribed in this study.
As discussed by \citeA{hourdin2017art}, models are often calibrated through a combination of parameter tuning and the selection of model forms to produce results over an acceptable tolerance.
Although unbounded prior distributions could in principle be assigned to the parameters, we argue that such choices could hinder from addressing compensation errors indefinitely.
Therefore, in this study, the parameter ranges are uniform distributions informed by a previous study \cite{baki2022determining}, and are used to define a reasonable and wide domain for the global sensitivity analysis.
We randomly sample from these distributions to conduct global sensitivity analysis.
Parameter $a$ is a coefficient for the effect of the surface buoyancy flux on virtual potential temperature at the PBL height $h$.
Parameter $b$ is a constant of proportionality for the effect of the countergradient mixing term.
Parameter $p$ is the profile-shape exponent in the formulation of eddy viscosity.
Parameter $d_1$ and $d_2$ are coefficients that formulate the thickness of the entrainment zone.
From the SFCLAY scheme, we do not consider parameters in the Monin–Obukhov stability functions, as prior studies report low sensitivity \cite{yang2017sensitivity}.
Instead, we focus on the aerodynamic roughness length $z_0$, which is reported to exert a dominant influence on surface-layer turbulence \cite{hahmann2020making}.
We note that $z_0$ varies with surface characteristics and is therefore not inherently constant.
In this study, however, we prescribe a range of $z_0$ values so that its influence can be compared consistently with that of the other PBL scheme parameters.
In this sense, it plays a role similar to that of parameter \textit{znt\_zf} in \citeA{baki2022determining}.
Similarly, future work could investigate the parameters within anemometric or morphometric surface roughness models that are used to predict across different surface morphologies \cite{macdonald1998improved,grimmond1999aerodynamic,yang2016exponential,altland2025distributed}.

\begin{table}[t]
\centering
\caption{Target parameters in the planetary boundary layer (PBL) and surface-layer (SFCLAY) schemes}
\label{tab:sl-pbl-parameters}
\begin{tabular}{lllll}
\hline\noalign{\smallskip}
Parameterization type & Scheme & Parameter & Standard value & Range \\
\hline\noalign{\smallskip}
\multirow{5}{*}{Planetary boundary layer}
& \multirow{5}{*}{YSU}
& $a$     & 6.8  & $\mathcal{U}(3.4,13.6)$ \\
& & $b$     & 6.8  & $\mathcal{U}(3.4,13.6)$ \\
& & $p$     & 2    & $\mathcal{U}(1,3)$ \\
& & $d_{1}$ & 0.02 & $\mathcal{U}(0.01,0.03)$ \\
& & $d_{2}$ & 0.05 & $\mathcal{U}(0.025,0.075)$ \\
Surface layer & Revised MM5
& $z_{0}$ & 0.1 & $\mathcal{U}(0.05,0.20)$ \\
\noalign{\smallskip}\hline
\end{tabular}
\end{table}

\subsubsection{Idealized planetary boundary layer simulation cases}
\label{sec:simulationcases}

\begin{figure}
    \centering
        \includegraphics[width=0.9\textwidth]{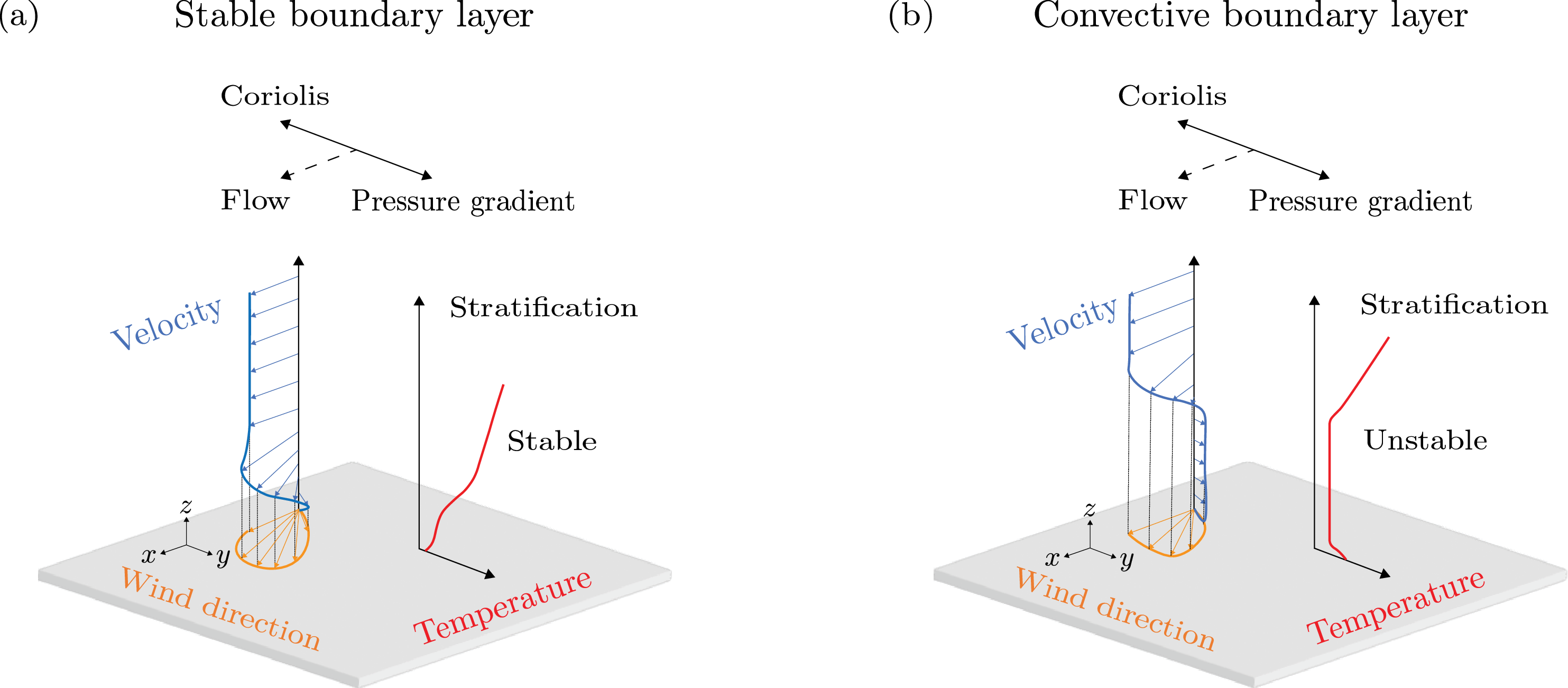}
    \caption{Two flow environments simulated in the WRF model: (a) stable boundary layer and (b) convective boundary layer. The geostrophic balance between the pressure-gradient force and the Coriolis force aloft is depicted for each flow environment.}
    \label{fig:abl_schematic}
\end{figure}

We consider idealized, horizontally homogeneous, dry PBL simulations under two different atmospheric stability regimes: a stable boundary layer and a convective boundary layer.
Periodic boundary conditions are used in the streamwise and cross-stream directions.
The computational setups for the two simulations are summarized in Table~\ref{tab:stabilitycases}, and the respective flow environments are visualized in Fig.~\ref{fig:abl_schematic}.
The setup for the stable boundary layer is informed by \citeA{beare2006intercomparison}, while the convective boundary layer is informed by \citeA{shin2016evaluation}.
Since the flow is horizontally homogeneous, the flow statistics are uniform in the horizontal ($x$,$y$) directions.
To preserve a clear definition of the averaging operator, model-output statistics are computed using temporal averaging only (no spatial averaging) throughout the study.

\begin{table}
\caption{Summary of computational setup for the two flow environments}
\label{tab:stabilitycases}
\centering
\begin{tabular}{lcc}
\hline\noalign{\smallskip}
Atmospheric stability & Stable & Convective \\
\noalign{\smallskip}\hline\noalign{\smallskip}
Domain height $z_{\mathrm{top}}~\mathrm{[m]}$ & $400$ & $3500$ \\
Grid points $N_z~\mathrm{[-]}$  & $64$ & $120$ \\
Geostrophic wind speed $u_G~\mathrm{[m~s^{-1}]}$ & $8$ & $10$ \\
Initial potential temperature profile $\theta(z)~\mathrm{[K]}$ & (Eq.~\ref{eq:sblpotentialtemperature}) & (Eq.~\ref{eq:cblpotentialtemperature}) \\
Free-atmosphere lapse rate $L~\mathrm{[K~m^{-1}]}$ & $0.01$ & $0.01$ \\
Latitude $\phi~\mathrm{[^\circ]}$ & $73$ & $43.3$ \\
Roughness length $z_0~\mathrm{[m]}$ & $0.1$ & $0.1$ \\
Surface cooling rate & $-0.25~\mathrm{K~h^{-1}}$ & $0.20~\mathrm{K~m~s^{-1}}$ \\
Spin-up time $\mathrm{[h]}$ & $8$ & $2$ \\
\noalign{\smallskip}\hline
\end{tabular}
\end{table}

\textbf{Stable boundary layer} The stable boundary layer is a poorly mixed layer that forms when the surface is cooler than the air above \cite{stull2012introduction}.
It is characterized by buoyancy-suppressed turbulent mixing and resulting turbulence intermittency, which show strong sensitivity to turbulence closures \cite{cuxart2006single, mahrt2010variability, mahrt2014stably}.
We adopt the moderately stable configuration from the GABLS intercomparison, which prescribes constant surface cooling to achieve statically stable conditions \cite{beare2006intercomparison}.
The initial potential temperature profile is
\begin{equation}
  \theta(z) =
  \begin{cases}
      265~\mathrm{K},            & 0 < z \le 100~\mathrm{m},\\
      265~\mathrm{K} + (z - 100~\mathrm{m})\, 0.01~\mathrm{K\,m^{-1}}, & z > 100~\mathrm{m}.
  \end{cases}
  \label{eq:sblpotentialtemperature}
\end{equation}
A surface cooling rate of $0.25~\mathrm{K}~\mathrm{h}^{-1}$ is prescribed.
A Rayleigh damping layer is used in the upper $100~\mathrm{m}$.
Time-averaged vertical profiles of model outputs are sampled after the 8th hour, when the flow reaches a statistical quasi-stationary regime.

\textbf{Convective boundary layer} The convective boundary layer is the buoyancy-driven mixed layer produced by surface heating and is characterized by strong thermals and well-mixed statistical properties of state variables \cite{moeng1994comparison, shin2013analysis}.
The setup from \citeA{shin2016evaluation} is adopted as the reference dry configuration. 
The initial potential temperature profile is
\begin{equation}
\label{eq:cblpotentialtemperature}
\theta(z)=
\begin{cases}
300~\mathrm{K}, 
& 0<z\le 925~\mathrm{m},\\[6pt]
300~\mathrm{K} + (z-925~\mathrm{m})\, 0.0536~\mathrm{K\,m^{-1}}, 
& 925<z\le 1075~\mathrm{m},\\[6pt]
308.05~\mathrm{K} + (z-1075~\mathrm{m})\, 0.003~\mathrm{K\,m^{-1}}, 
& z>1075~\mathrm{m}.
\end{cases}
\end{equation}
A Rayleigh damping layer is used in the upper $1000~\mathrm{m}$.
We sample time-averaged vertical profiles of model outputs after $12 \tau$, when the flow reaches a statistical quasi-stationary regime, where $\tau$ is the estimated large eddy turnover time at approximately $600~\mathrm{s}$ \cite{shin2013analysis}.

\subsubsection{Synthetic observations}
\label{sec:syntheticobservations}

\begin{figure}
    \centering
        \includegraphics[width=0.85\textwidth]{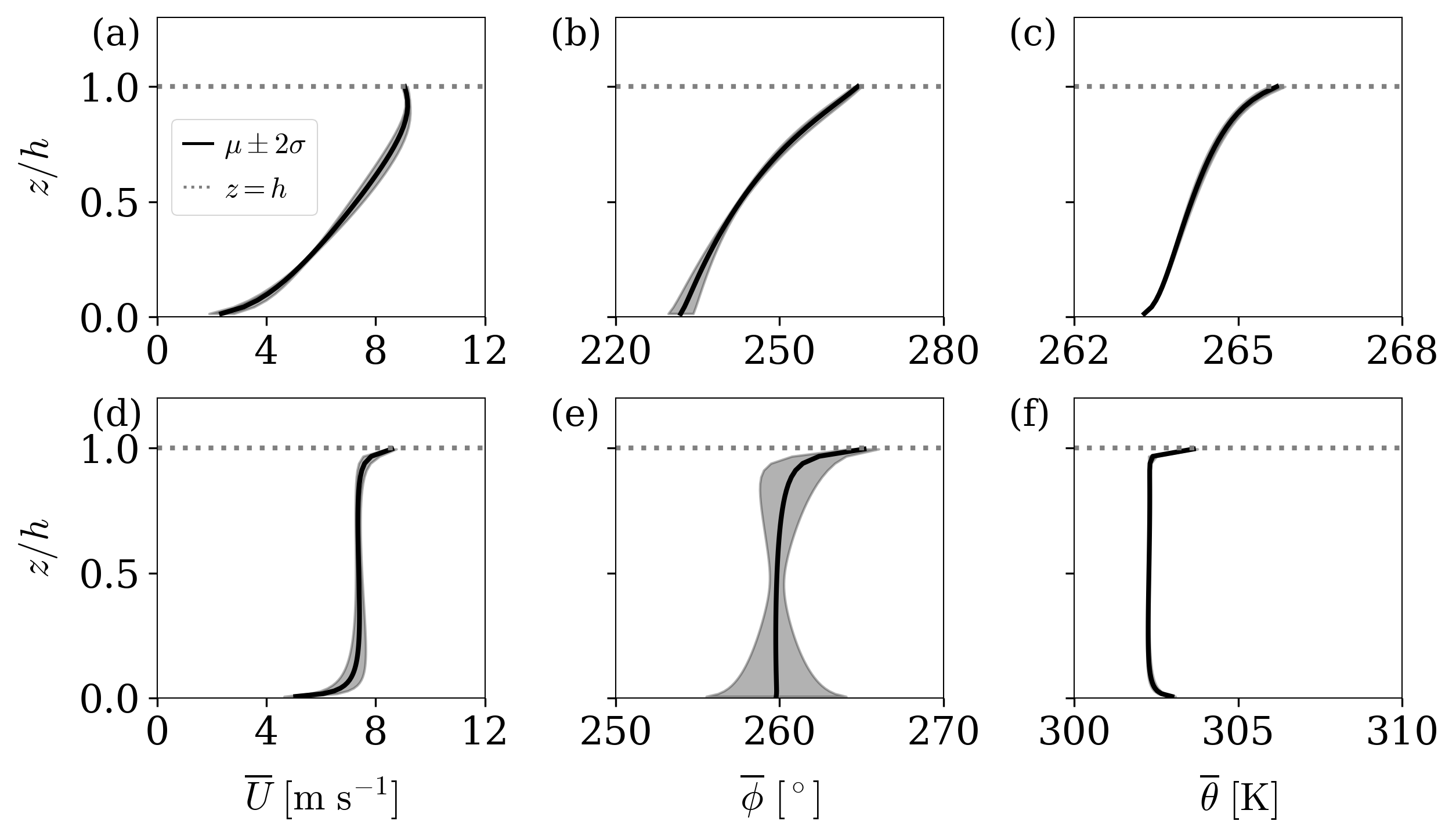}
    \caption{Synthetic observations generated by the coupled YSU PBL-revised MM5 scheme with standard parameters using different initial-condition realizations.
    Top row shows the stable boundary layer while bottom row shows the convective boundary layer for 30-minute-averaged output quantities of interest: (a,d) wind speed, (b,e) wind direction, and (c,f) potential temperature. 
    The vertical height is normalized by the ensemble-mean PBL height $h$ (dotted line).
    The ensemble spread ($\mu \pm 2\sigma$) represents internal variability arising from varied initial conditions via random-seed perturbations.
    }
    \label{fig:synthetic_observations}
\end{figure}

In this study, we conduct a perfect-model test in which we infer parameters from observations that are generated by the PBL scheme itself using prescribed \textit{truth} parameter values.
Because the system is chaotic, changes in the initial conditions introduce internal variability into the observations.
We refer to these as \emph{synthetic} observations, which are useful in proof-of-concept studies for the two following reasons: (i) successful recovery of the prescribed truth parameters provides a measure of how informative the observations are for parameter calibration, and (ii) potential model-form error can be removed from the inverse problem formulation in conducting parameter calibration.
Although we do not consider an explicit treatment of model-data discrepancy in this study, we view it as a critical component and leave it for future work.
The synthetic observations are generated by fixing parameters at their standard values and varying the initial conditions using different random seeds that introduce independent, uniformly distributed perturbations drawn from $\mathcal{U}(-0.05, .05)\,\mathrm{K}$ to the initial potential temperature profile in the lowest four vertical grid levels.
The resulting statistics reflect the internal variability arising from the chaotic dynamics of the PBL system.
Figure~\ref{fig:synthetic_observations} presents time-averaged vertical profiles of wind speed, wind direction, and potential temperature using the coupled YSU PBL-revised MM5 scheme for the stable and convective cases detailed in Table~\ref{tab:stabilitycases}.
Because PBL schemes parameterize turbulence within the PBL, we focus on output statistics up to the PBL height $h$, which is defined as the vertical location of the largest increase in potential temperature \cite{sullivan1998structure}.
These observations drive the Bayesian inference of target parameters via an observed sample and an observational noise covariance matrix.

\FloatBarrier

\section{Results}
\label{sec:results}

In the following section, we present results on (1) global sensitivity analysis of the coupled PBL-SFCLAY scheme parameters and (2) the observation selection framework in connection with the inverse UQ of the parameters.

\subsection{Global sensitivity analysis}
\label{sec:variancebasedsensitivityanalysisresults}

\begin{table}
\caption{Parametric study for global sensitivity analysis}
\label{tab:parametricstudygsa}
\centering
\begin{tabular}{lc}
\hline\noalign{\smallskip}
\textbf{Factor} & \textbf{Options} \\ 
\noalign{\smallskip}\hline\noalign{\smallskip}
Quantity of interest & Wind speed, wind direction, potential temperature \\ 
Time-averaging window [min] & 5, 10, 30, 60 \\
Atmospheric stability regime & Stable, convective \\ 
Vertical location [-] & $0 \le z/h \le 1$ (regime-dependent $h$) \\
\noalign{\smallskip}\hline
\end{tabular}
\end{table}

\begin{figure}
    \centering
        \includegraphics[width=0.85\textwidth]{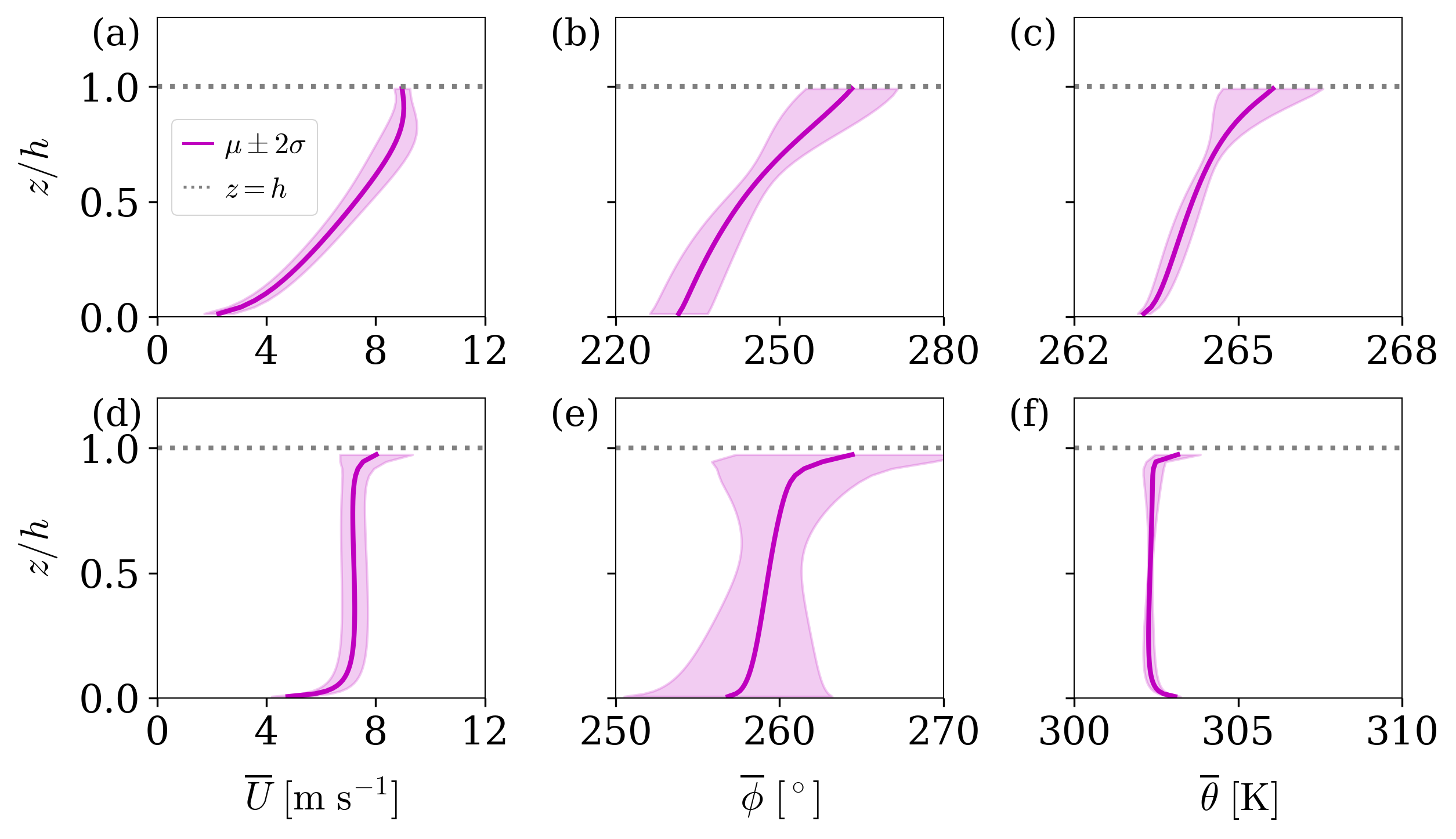}
    \caption{Emulator training data ($\mu\pm2\sigma$) for model outputs by the coupled YSU PBL-revised MM5 scheme, based on 200 model evaluations sampled from the parameter distributions in Table~\ref{tab:sl-pbl-parameters}. 
    The top row shows results of the stable boundary layer, while the bottom row shows results of the convective boundary layer.
    Panels show 30-minute-averaged vertical profiles of (a,d) wind speed, (b,e) wind direction, and (c,f) potential temperature.
    }
    \label{fig:emulator_training_outputs}
\end{figure}

Global sensitivity analysis is performed using GP emulators trained to map the coupled YSU PBL-revised MM5 scheme parameters $\boldsymbol{X} \in \mathbb{R}^p$ to vertical profiles of model outputs $f(\boldsymbol{X} )\in \mathbb{R}^q$, where $q$ is the model-output dimension.
Note that here we only consider the dependence of the model outputs on the vertical location $z$ because the flow is horizontally homogeneous, but this can be naturally formulated at arbitrary positions in space in a general setting with heterogeneity in the statistics in the horizontal directions.
First, we emphasize that this statistical analysis would not be possible without computationally efficient emulators.
For the $p=6$ parameters in Table~\ref{tab:sl-pbl-parameters} and an empirically determined sample size $N=50\,000$, computing the main- and interaction-effect Sobol’ sensitivity indices for a single model output would require $N(2p + 2)=700\,000$ evaluations of the WRF model.
Instead, the emulator is trained using outputs from 200 WRF simulations generated with different parameter combinations (Fig.~\ref{fig:emulator_training_outputs}).
We find that a training set of 200 input–output pairs (WRF model evaluations) achieves convergence of the surrogate-based estimators, consistent with a previous finding \cite{baki2022determining}.
The sensitivity of the emulator-based Sobol’ sensitivity indices to both the number of WRF model evaluations used for training and the emulator sample size $N$ becomes negligible once these quantities are sufficiently large (see \ref{sec:appendix1}).

For the modeled representation of the turbulent PBL, we expect parameter sensitivities of the PBL and SFCLAY schemes to vary across space and time, across output quantities, and across flow regimes.
Therefore, the sensitivity analysis examines how parameter sensitivity depends on four factors: quantity of interest, time-averaging window, atmospheric stability, and vertical location.
The factors and the corresponding options used in the parametric study are summarized in Table~\ref{tab:parametricstudygsa}.
First, because atmospheric stability strongly affects parameter sensitivities, results are shown separately for stable and convective boundary layers at a fixed 30-minute averaging window.
Because atmospheric stability plays a dominant role in PBL processes, representative stability regimes may be leveraged to provide generalizable insights into parameter sensitivity behavior \cite{yang2017sensitivity}.
Although not shown here, we confirm that sensitivity results are robust to moderate variations in surface and mesoscale forcings within each stability regime (see \ref{sec:appendix2}).

\begin{figure}
    \centering
        \includegraphics[width=0.85\textwidth]{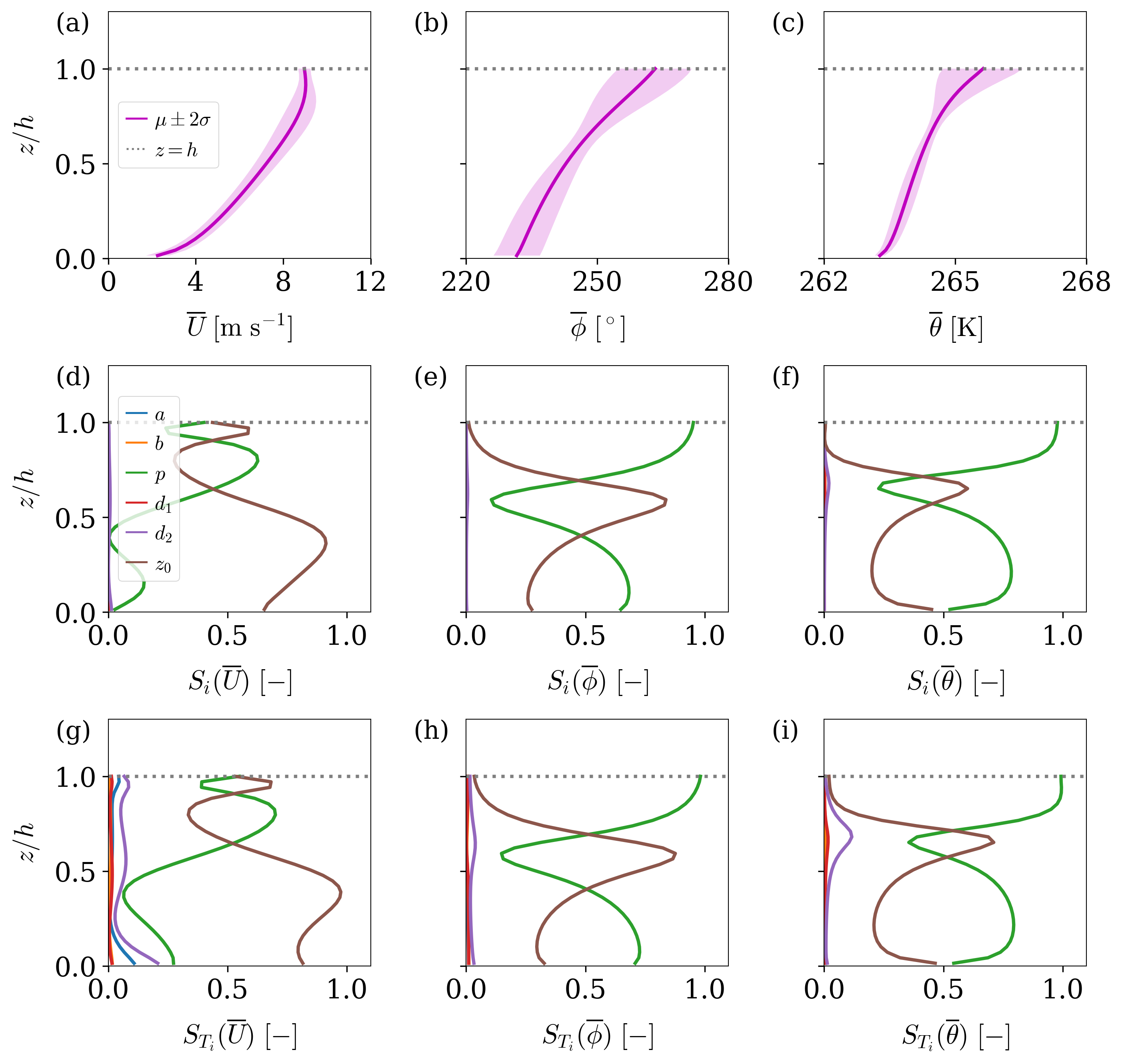}
    \caption{30-minute-averaged profiles of (a) wind speed, (b) wind direction, and (c) potential temperature in the stable boundary layer: (d–f) main-effect and (g–i) total-effect sensitivity indices of parameters in the coupled YSU PBL-revised MM5 scheme}
    \label{fig:sobol_indices_ysu_stable_30min_diff_qois}
\end{figure}

\begin{figure}
    \centering
        \includegraphics[width=0.85\textwidth]{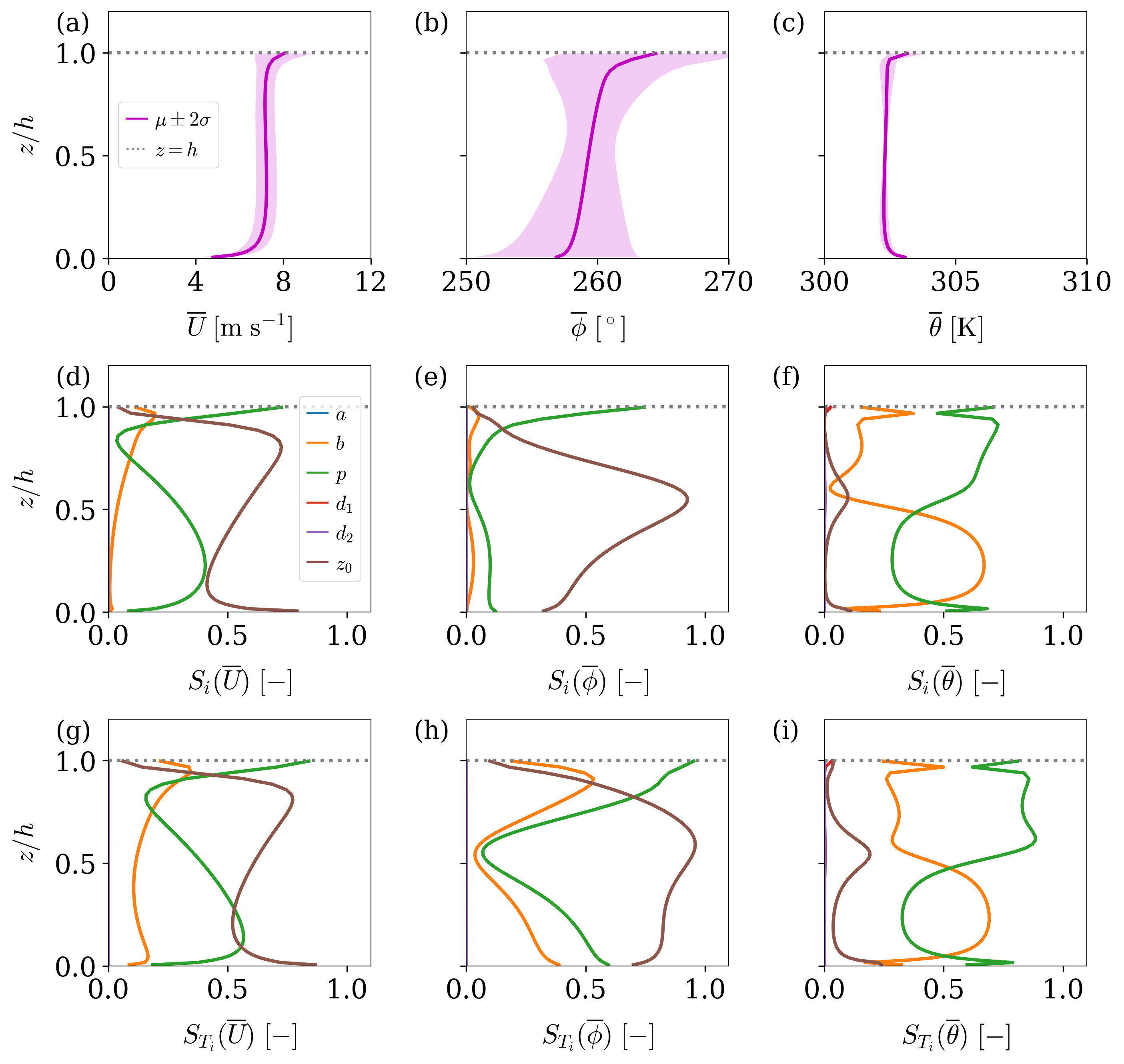}
    \caption{30-minute-averaged profiles of (a) wind speed, (b) wind direction, and (c) potential temperature in the convective boundary layer: (d–f) main-effect and (g–i) total-effect sensitivity indices of parameters in the coupled YSU PBL-revised MM5 scheme}
    \label{fig:sobol_indices_ysu_convective_30min_diff_qois}
\end{figure}

In the stable regime, two parameters, the profile-shape exponent $p$ and roughness length $z_0$, dictate the vertical profiles of wind speed, wind direction, and potential temperature.
Figure~\ref{fig:sobol_indices_ysu_stable_30min_diff_qois} shows the parameter sensitivities in the coupled YSU PBL-revised MM5 scheme for 30-minute-averaged statistics of the stable boundary layer: the first row shows the three output quantities of interest, followed by the corresponding main-effect and total-effect sensitivity indices.
Heights in all profiles are normalized by PBL height $h$ (dotted line).
First, the contribution of $z_0$ is large near the surface, accounting for up to $91\%$ in independent contribution to wind-speed variance, as surface roughness controls the imposed boundary conditions of surface momentum and heat fluxes.
In contrast, the contribution of $p$ increases with height and dominates near the peak of the low-level jet, independently contributing up to $63\%$ of the total wind-speed variance.
The main-effect sensitivity indices (Fig.~\ref{fig:sobol_indices_ysu_stable_30min_diff_qois}d-f) are quantitatively similar to the total-effect counterparts (Fig.~\ref{fig:sobol_indices_ysu_stable_30min_diff_qois}g-i), which indicates that parameters contribute primarily independently to the model-output variance.
Consistent with the modeled representation of flow physics by the YSU PBL scheme \cite{troen1986simple,hong1996nonlocal,hong2006new}, the other parameters are inactive in the stable regime.

In the convective regime (Fig.~\ref{fig:sobol_indices_ysu_convective_30min_diff_qois}), three parameters, the profile-shape exponent $p$, countergradient mixing constant $b$, and roughness length $z_0$, dominantly shape the wind speed, wind direction and potential temperature.
In contrast to those of the stable regime, main-effect indices reveal that the countergradient mixing constant $b$ (orange) is non-negligible in the convective PBL, especially in shaping the potential temperature profile.
Its contribution is particularly strong in the lower half of the mixed layer for potential temperature, where it accounts for approximately $67\%$ in the main-effect contribution.
This is because parameter $b$ plays a key role in the countergradient term for heat flux through its coupling with the surface flux and PBL height.
On the other hand, for wind speed in the convective regime, the contribution of the roughness length $z_0$ dominates throughout the surface layer, decreases to approximately $41\%$ in main-effect contribution, and subsequently increases to $73\%$ in the upper mixed layer.
Near the PBL height, parameter $p$, which directly controls the eddy-viscosity profile, independently contributes up to $73\%$ of the variance in wind speed.
Because the YSU PBL scheme does not have a separate buoyancy-dependent parameter that explicitly accounts for buoyancy suppression at the temperature inversion of the free atmosphere, this effect is implicitly represented through parameter $p$.
The resulting high contribution of $p$ near the PBL height in both stable and convective regimes clearly demonstrates the opportunity and, perhaps, the necessity to model $p$ as a function of atmospheric stability in future work.
In this way, parameter sensitivity can help identify areas where physics parameterizations may require further development to represent physical mechanisms more accurately.

\begin{figure}
    \centering
        \includegraphics[width=1.0\textwidth]{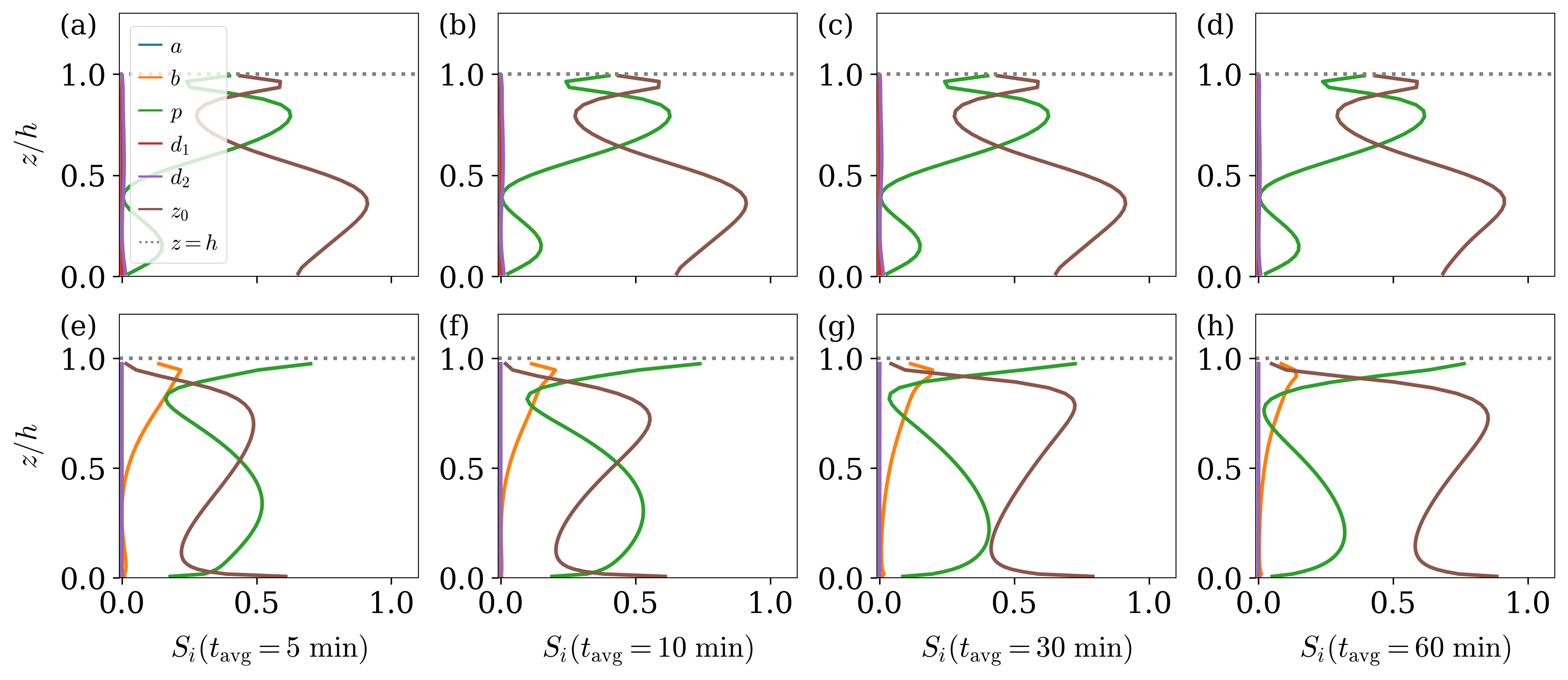}
    \caption{
    Main-effect sensitivity indices of the coupled YSU PBL-revised MM5 scheme parameters for modeled wind speed for centered time-averaging windows, shown for stable (top row) and convective (bottom row) regimes. Panels (a,e), (b,f), (c,g), and (d,h) correspond to the four averaging windows.
    }
    \label{fig:first_indices_ysu_wsph_diff_tavgs}
\end{figure}

We next investigate how temporal averaging in the model outputs influences parameter contributions to the variance of the model output.
Temporal averaging is important in the turbulent PBL because the flow is inherently transient, with dominant energy contributions occurring across a range of frequency regimes--from mesoscale forcing to turbulence scales.
This necessitates adequate averaging in both model outputs and observations so that they can be compared meaningfully for parameter calibration in turbulence parameterizations.
In the chaotic PBL system, the adequacy of this averaging is related to the timescales of the largest turbulent eddies in the flow, which is affected by the atmospheric stability regime \cite{metzger2008time, mahrt2020non}.
As a result, we observe that the main-effect sensitivity indices of the coupled YSU PBL-revised MM5 scheme parameters for wind speed differ significantly between stable and convective regimes across the four centered time-averaging windows considered (Fig.~\ref{fig:first_indices_ysu_wsph_diff_tavgs}).
In the stable regime (Fig.~\ref{fig:first_indices_ysu_wsph_diff_tavgs}a-d), negligible changes in the relative balance of parameter contributions are observed across the averaging windows.
This suggests that the largest turbulent fluctuations in the flow occur on timescales shorter than the shortest averaging window of 5 minutes.
By contrast, in the convective regime (Fig.~\ref{fig:first_indices_ysu_wsph_diff_tavgs}e-h), as the averaging window increases, we observe that the parameter sensitivity balance between primarily profile-shape exponent $p$ and roughness length $z_0$ shift.
The contribution of $z_0$ within the mixed layer for 60-minute averaging increases by up to $38\%$ relative to 5-minute averaging, while the relative contribution of $p$ correspondingly decreases with longer temporal averaging.
This shift indicates that as the averaging window averages over more turbulent fluctuations, the mean flow becomes increasingly controlled by surface drag via the roughness length.

\begin{figure}
    \centering
        \includegraphics[width=1.0\textwidth]{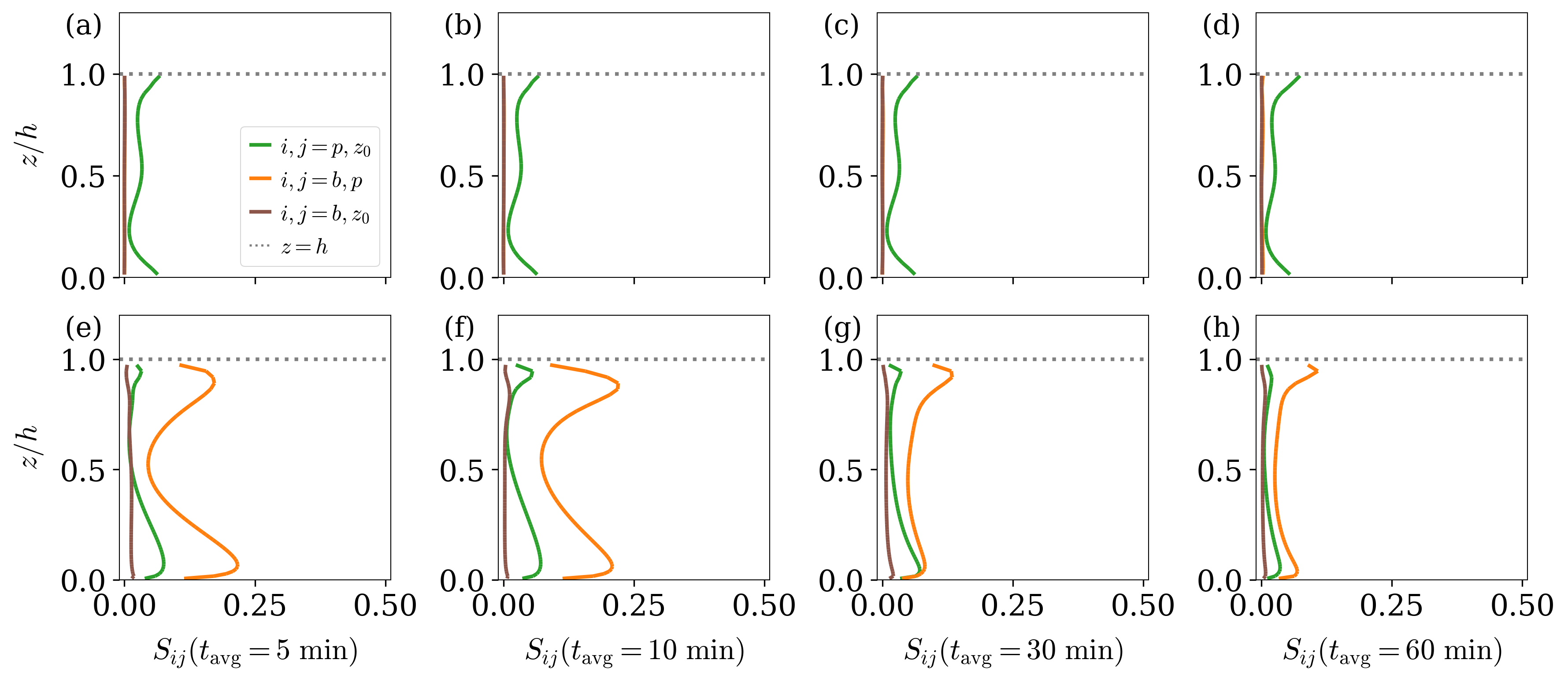}
    \caption{Interaction-effect sensitivity indices of the coupled YSU PBL-revised MM5 scheme parameters for modeled wind speed for centered time-averaging windows, shown for stable (top row) and convective (bottom row) regimes. 
    Panels (a,e), (b,f), (c,g), and (d,h) correspond to the four averaging windows.
    }
    \label{fig:second_indices_ysu_wsph_diff_tavgs}
\end{figure}

The relationship between $p$ and $z_0$ is further explored in the second-order interaction effects of parameters in the PBL and SFCLAY schemes to modeled wind speed in Fig.~\ref{fig:second_indices_ysu_wsph_diff_tavgs}.
In the stable case (Fig.~\ref{fig:second_indices_ysu_wsph_diff_tavgs}a-d), we observe an approximately $7\%$ interaction contribution to wind speed variance by the profile-shape exponent $p$ and roughness length $z_0$ in the stable regime.
This interaction contribution has local peaks at the surface and the PBL height $h$ (dotted line).
The magnitude of this interaction is invariant to temporal averaging, implying an irreducible dependence between the two parameters that is fundamental to the modeled wind speed profile in stable stratification.
This result reinforces the position that PBL and SFCLAY schemes should be considered jointly in parameterization development and calibration \cite{baklanov2011nature}.
Because the counter-gradient coefficient $b$ is inactive in the stable regime, both its interactions with $p$ and $z_0$ are zero.
In contrast, the bottom row of Fig~\ref{fig:second_indices_ysu_wsph_diff_tavgs} shows the interaction between $p$ and $z_0$ for the convective boundary layer, which decays with longer averaging.
In the convective regime, the interaction between parameters $b$ and $p$ is active and strongest at the shortest averaging window of 5 minutes.
Although this also decays with longer averaging, we observe an approximately $9\%$ interaction contribution to wind speed variance at 60-minute averaging, which is non-negligible.
Finally, there is an interaction between $b$ and $z_0$, which is almost zero.
These non-negligible, nonlinear parameter interaction contributions to the model-output variance, which arise even in temporally averaged statistics, can potentially lead to identifiability issues in parameter calibration \cite{lam2022practical}.
That is, due to interactions between parameters, multiple distinct parameter values can produce similarly acceptable model predictions, a well-known phenomenon known as \textit{equifinality} \cite{regayre2018aerosol}.
We revisit the importance of temporal averaging in the context of Bayesian inference for parameters in the following section.

To summarize, the results of emulator-aided global sensitivity analysis elucidate the sensitivity of the model to parameters in coupled physics parameterizations across multiple factors important to characterizing model predictions of the chaotic dynamical system.
Accurate statistical analysis of the forward mapping from parameter space to model-output space is made possible by lightweight, modular ML emulators.
The analysis reveals which parameters dominate, where they act across the model-output space, and how they interact across different timescales.
It also reveals that some parameter interactions persist under long temporal averaging; these interactions should be examined further to determine whether they reflect, or in some cases reveal, underlying physical mechanisms in the flow.
This suggests a direction for future work to apply this method to investigate nonlinear interactions among parameters and across multiple coupled physics parameterizations.
Although the benefits of the sensitivity analysis are clear, the forward propagation of uncertainty is not a substitute for inverse UQ, since characterizing physically meaningful credible bounds for parameters requires Bayesian inversion using observations.

\FloatBarrier

\subsection{Observation selection for Bayesian inversion}
\label{sec:observationselectionforbayesianinversion}

The selection of observations in a chaotic dynamical system is particularly relevant to assimilating field observations for parameterization development in ESMs.
Previous results in Sect.~\ref{sec:variancebasedsensitivityanalysisresults} showed that the parameter sensitivity varies significantly with factors that characterize model predictions.
In the subsequent results, we explore how each of these factors, when used to construct observations that drive inverse UQ, influences the inferred parameter uncertainty.
In doing so, we explore how parameter sensitivity can be leveraged using the diagnostic measures introduced in Sect.~\ref{sec:observationselectionframework} to calibrate and reduce uncertainty in PBL and SFCLAY scheme parameters.
As a demonstration of the potential in this study, we use synthetic observations (see Sect.~\ref{sec:syntheticobservations}) generated using the same WRF model such that model error is eliminated from the inversion process.
Here, we consider the number of spatial locations to observe in the context of meteorological masts, which are a practical source of observations in the PBL.
Therefore, we fix the number of observed spatial locations to five--within the range commonly used in operational meteorological masts \cite{ramon_tall_2020}.
Fixing the number of observed locations eliminates the confounding variable of data volume.
These locations are uniformly allocated across the observation space, which in this application, is up to the PBL height $h$.
Observations from these locations are used to drive emulator-aided Bayesian inversion via Metropolis–Hastings MCMC sampling, using $10\,000$ burn-in iterations and $200\,000$ posterior samples \cite{shin2026accelerated}.

\subsubsection{Influence of observed atmospheric stability}
\label{sec:influenceofobservedatmosphericstability}

\begin{figure}
    \centering
        \includegraphics[width=0.85\textwidth]{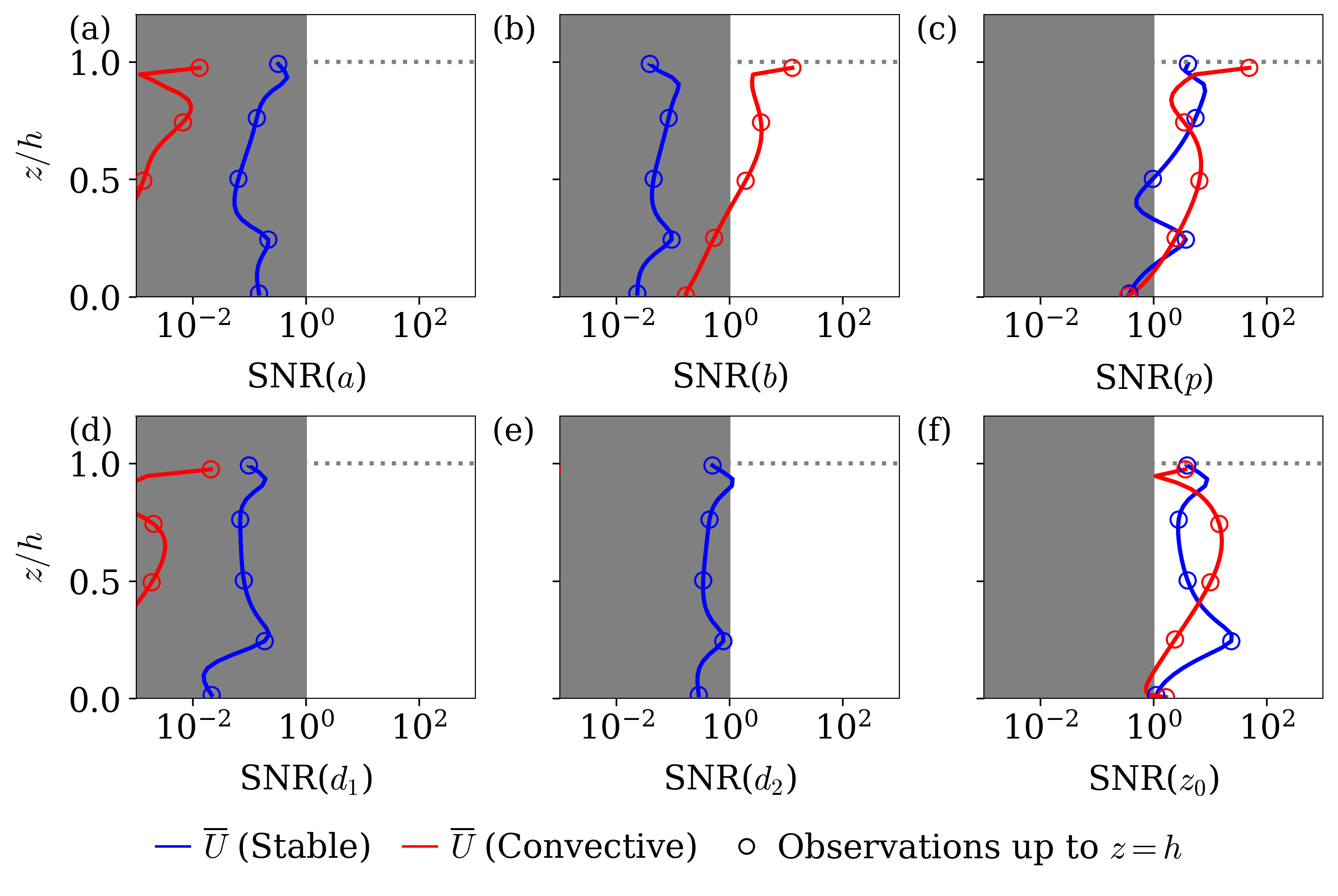}
    \caption{
    Signal-to-noise (SNR) ratios computed for the coupled YSU PBL-revised MM5 scheme parameters based on 30-minute-averaged wind speed in the stable (blue) and convective (red) regimes.
    Circle markers correspond to the observed vertical locations up to PBL height $h$
    }
    \label{fig:snr_ysu_wsph_30min_diff_stabilities}
\end{figure}

We step through the observation selection workflow as we explore how the inferred parameter uncertainty is influenced by the aforementioned four factors: observed quantity of interest, atmospheric stability during observation, observation averaging time length, and the spatial location of the observations.
First, we investigate the influence of the observed atmospheric stability.
To isolate this influence, we fix the three other factors to 30-minute-averaged wind speed observed in five uniformly distributed locations up to the PBL height.
We compute the first diagnostic measure, the signal-to-noise ratio, for the coupled YSU PBL-revised MM5 scheme parameters based on 30-minute-averaged wind speed in the stable (blue) and convective (red) stability regimes (Fig.~\ref{fig:snr_ysu_wsph_30min_diff_stabilities}).
The signal-to-noise ratio evaluates whether the total parameter signal is larger than observational noise across the observation space.
Vertical profiles of this ratio for each parameter compares the total dimensionalized variance contribution by a parameter to the observational noise variance at each spatial location.
We visualize $\mathrm{SNR}(i) \ge 1$ as the non-shaded region an $\mathrm{SNR}(i) < 1$ as the gray shaded region.
Therefore, parameters whose effects exceed observational noise lie within the non-shaded region, whereas those that do not exceed it lie within the shaded region.

From the signal-to-noise ratio of each parameter, two trends are clear about the sensitivity of 30-minute-averaged wind speed to the coupled YSU PBL-revised MM5 scheme parameters regardless of atmospheric stability.
First, the dimensionalized contributions of parameters $p$ and $z_0$ to the wind speed exceed observational noise across most of the PBL.
Second, the signals of parameters $a$, $d_1$, and $d_2$ are not visible over the observational noise, a situation referred to as \textit{practical non-identifiability} \cite{lam2022practical}.
However, there is a clear influence of atmospheric stability to the visibility of the countergradient term constant $b$ as we observe that $b$ is observable in the upper mixed layer of the convective regime but not in the stable regime.
This is consistent with our understanding of the parameterization for stably stratified flows, since countergradient term parameters, $a$ and $b$, and the entrainment zone-specific parameters, $d_1$ and $d_2$, are not active in the stable regime.
In this way, prior to any inference process, we are able discern for the specified model output that profile-shape exponent $p$ and roughness length $z_0$ are observable above observational noise in both stability regimes and that $b$ is observable only in the convective regime.
This comparison of parameter signals with observational noise is possible because parameter sensitivity is projected onto the observation space.

\begin{figure}
    \centering
        \includegraphics[width=0.85\textwidth]{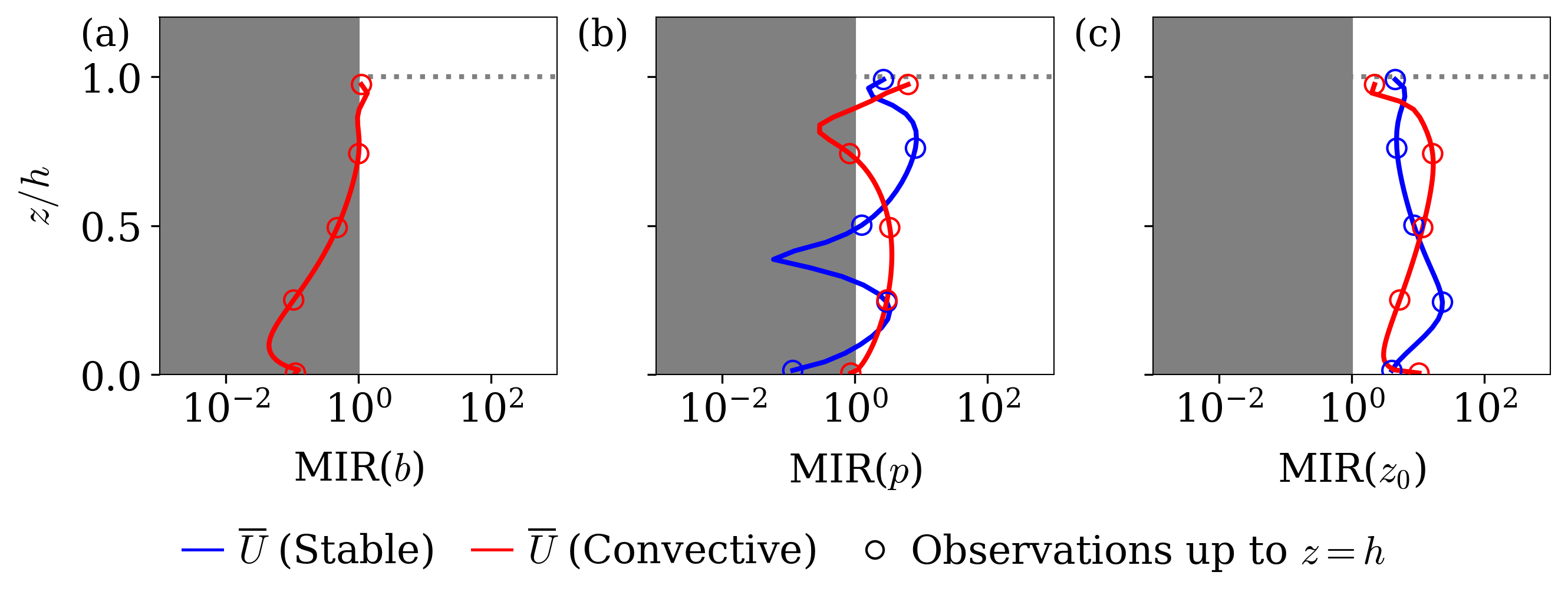}
    \caption{
    Main-to-interaction (MIR) ratios computed for the coupled YSU PBL-revised MM5 scheme parameters retained after filtering based on the corresponding signal-to-noise ratios, based on 30-minute-averaged wind speed in the stable (blue) and convective (red) regimes.
    Circle markers correspond to the observed vertical locations up to PBL height $h$
    }
    \label{fig:mir_ysu_wsph_30min_diff_stabilities}
\end{figure}

Although parameters may be calibrated in regimes where interaction effects with other parameters are strong, this calibration is then strictly dependent upon the specific values of other parameters.
We compute the main-to-interaction ratio to discern regions in the observation space where the independent contribution by a parameter to the model-output variance is larger than its interaction contributions.
This is performed for parameters that are retained after filtering based on the signal-to-noise ratio.
The main-to-interaction ratios of parameters $b$, $p$, and $z_0$ (Fig.~\ref{fig:mir_ysu_wsph_30min_diff_stabilities}) demonstrate that the uniformly distributed observed spatial locations for wind speed are clearly dominated by independent contributions for parameters $p$ and $z_0$ for both stability regimes.
For parameter $b$, the main-effect and interaction-effect contributions are comparable, but the ratio slightly exceeds unity near the PBL height.
Based on our criterion for this diagnostic measure, this indicates that $b$ is also practically identifiable from observations of 30-minute-averaged wind speed in the convective regime.
As described earlier, the signal-to-noise and main-to-interaction ratios together diagnose the practically identifiability of a parameter, which is the sufficiency condition for parameter calibration.

\begin{figure}
    \centering
        \includegraphics[width=1.0\textwidth]{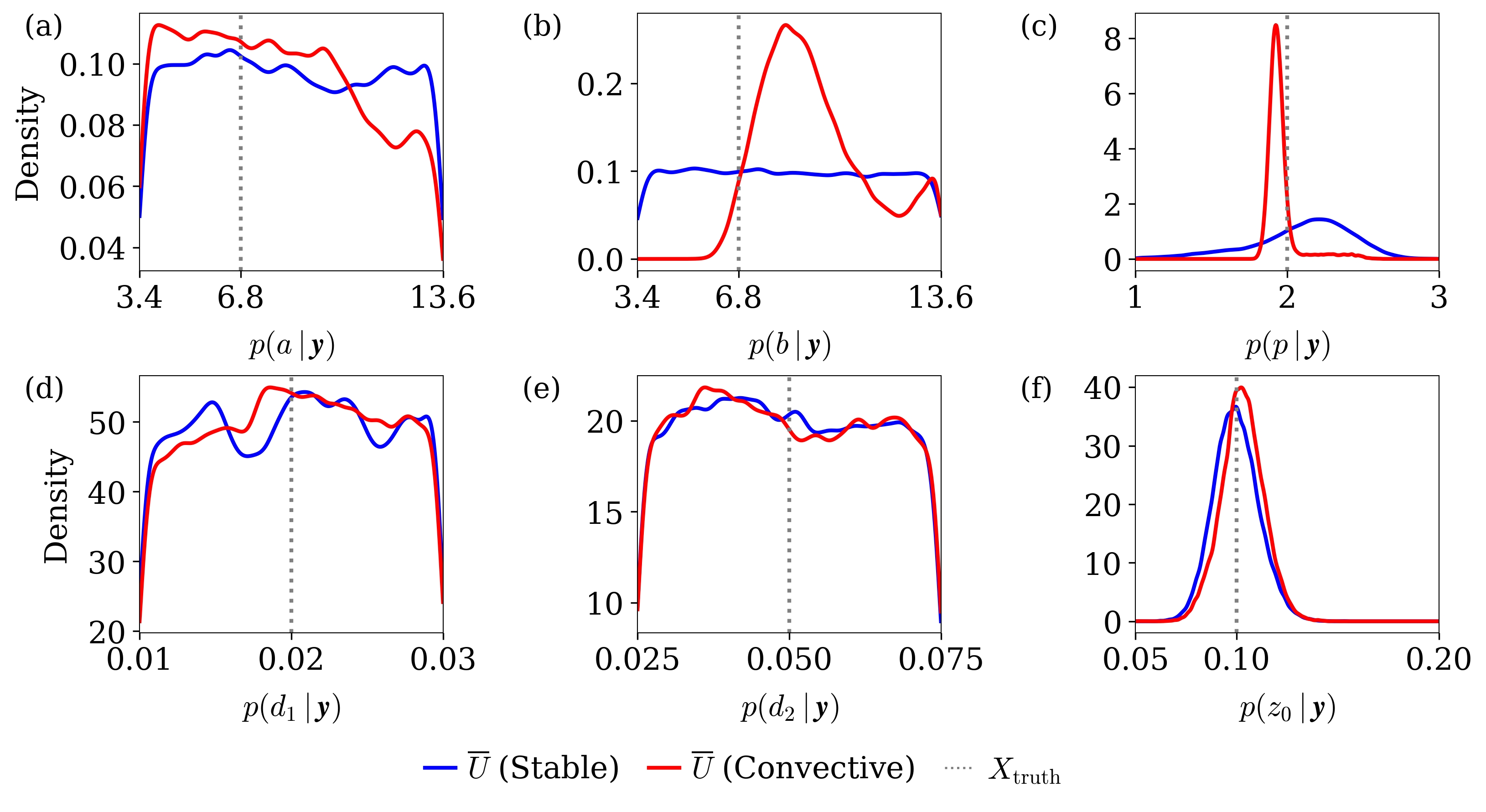}
    \caption{Marginal posterior distributions of the coupled YSU PBL-revised MM5 scheme parameters inferred from 30-minute-averaged wind speed in the stable (blue) and convective (red) regimes. The truth parameter value is indicated by the dotted line.}
    \label{fig:posterior_ysu_wsph_30min_diff_stabilities}
\end{figure}

Having examined the diagnostic measures of the parameter sensitivity we derived from forward uncertainty propagation, we now reverse the perspective and ask what observations from the two different atmospheric stability regimes reveal about the parameters.
By MCMC sampling from the emulators, we infer the posterior distributions of the coupled YSU PBL-revised MM5 scheme parameters from 30-minute-averaged wind speed observations for the two atmospheric stability regimes.
Figure~\ref{fig:posterior_ysu_wsph_30min_diff_stabilities} compares the inferred marginal posterior distributions of parameters for the stable (red) and convective (blue) regimes.
For the stable regime, parameters $p$ and $z_0$ show narrowed credible intervals that capture the truth parameter values (dotted line) close to the posterior mode, whereas the remaining four parameters show no reduction with respect to the range of the respective prior distributions.
In contrast, in the convective regime, we observe that additionally, parameter $b$ shows narrowed credible intervals close to the truth parameter value of $b$.
These results corroborate our analysis performed prior to the inference process using the signal-to-noise and main-to-interaction ratios.
This simple example highlights the effectiveness of the diagnostic measures in the observation selection framework as a tool for selecting observations to learn targeted parameters in ESM parameterizations.
Based on a comparison of the parameter signal with observational noise and its sensitivity contributions by independent and interaction effects, we can diagnose which parameters and where they are expected to be practically identifiable in the observation space.

\subsubsection{Influence of temporal averaging of observations}
\label{sec:influenceoftemporalaveragingofobservations}

Using our observation selection framework, we next investigate the influence of temporal averaging of observations to the inferred parameter uncertainty.
Temporal averaging of observations on the Bayesian inversion of ESM parameterizations has useful implications for measurement acquisition campaigns, data assimilation strategies, and parameter calibration workflows for the inherently transient, chaotic PBL system \cite{metzger2008time, calaf2023boundary}.
In the chaotic PBL system, mesoscale forcing timescales dictate the large-scale transience, while turbulence timescales dictate the turbulent energetics of the flow.
Therefore, to calibrate turbulence parameterizations in non-stationary dynamics without also performing state estimation, the time-average window must necessarily be longer than the turbulent timescales but possibly shorter than mesoscale forcing timescales if the goal is to preserve dependence on atmospheric phenomena with diurnal variability.
To isolate the influence of temporal averaging in our analysis, we fix the three other factors to convective-regime wind speed observed in the previously specified five vertical locations up to the PBL height.
We investigate for the aforementioned four time-averaging windows ($5$, $10$, $30$, $60$ minutes), which span the relevant averaging timescales reported to yield statistically stationary turbulence statistics in the convective regime \cite{metzger2008time}.
In results not shown, we observe that temporal averaging has negligible effect on the stable regime because our smallest temporal window of $5$ minutes averages over the integral timescales of the flow.

\begin{figure}
    \centering
        \includegraphics[width=0.85\textwidth]{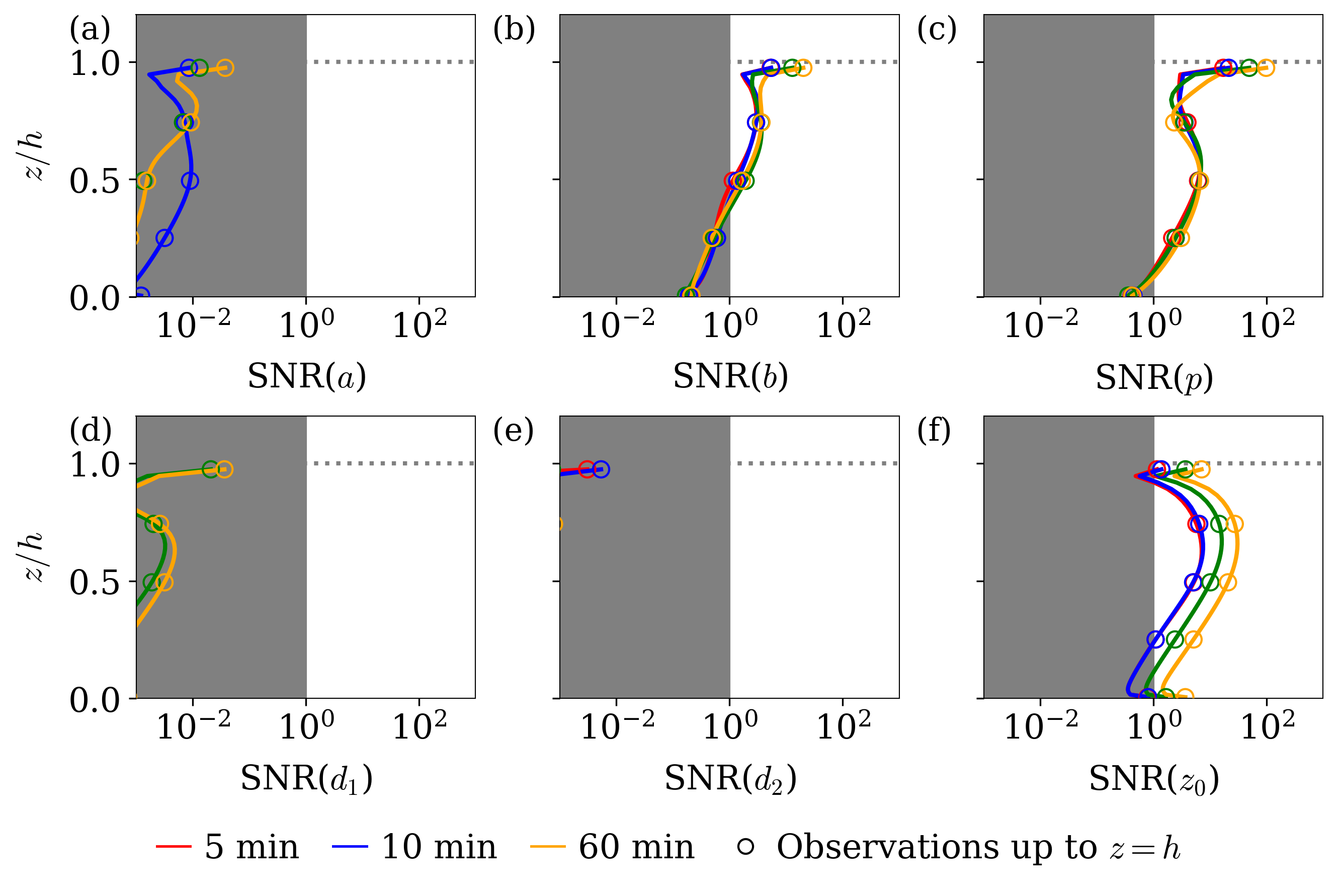}
    \caption{
    Signal-to-noise (SNR) ratios are computed for the coupled YSU PBL-revised MM5 scheme parameters based on wind speed of different time-averaging windows in the convective regime: 5 minute (red), 10 minute (blue), 30 minute (green), and 60 minute (orange).
    Circle markers correspond to the observed vertical locations up to PBL height $h$
    }
    \label{fig:snr_ysu_wsph_convective_diff_tavgs}
\end{figure}

\begin{figure}
    \centering
        \includegraphics[width=0.85\textwidth]{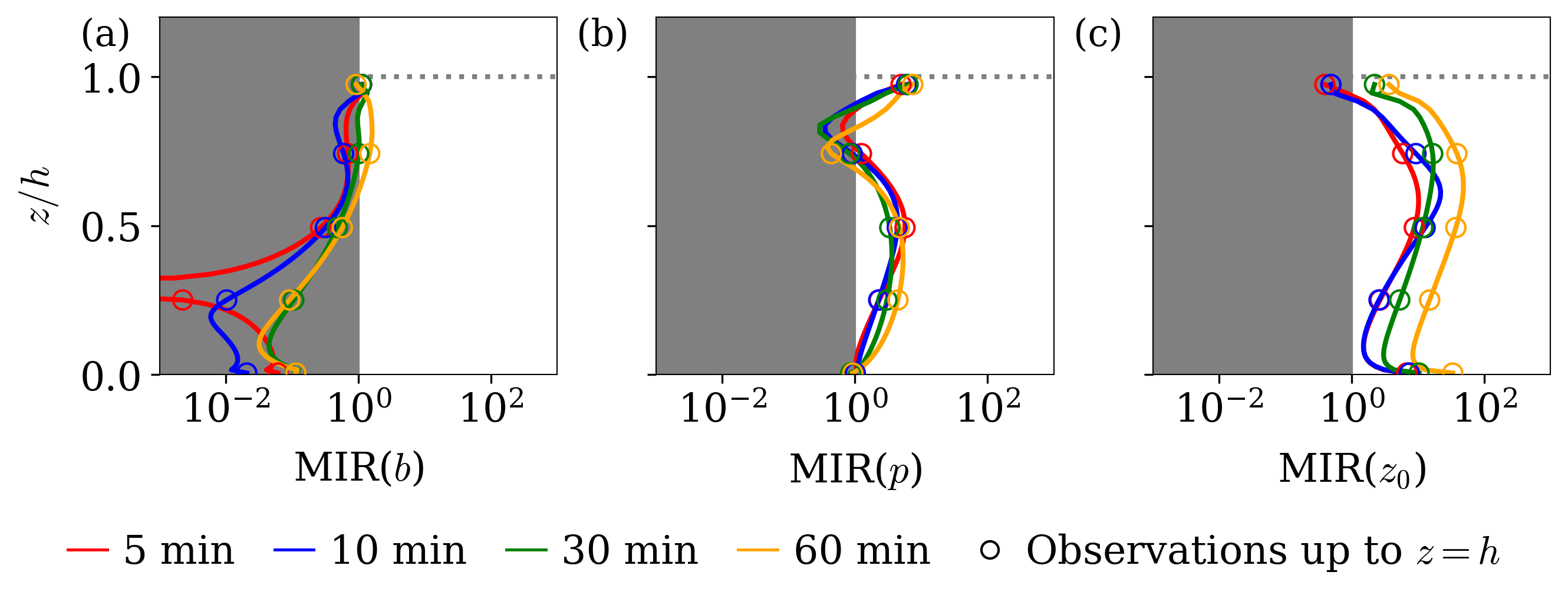}
    \caption{
    Main-to-interaction (MIR) ratios are computed for the coupled YSU PBL-revised MM5 scheme parameters retained after filtering based on the corresponding signal-to-noise ratios, based on wind speed of different time-averaging windows in the convective regime: 5 minute (red), 10 minute (blue), 30 minute (green), and 60 minute (orange).
    Circle markers correspond to the observed vertical locations up to PBL height $h$
    }
    \label{fig:mir_ysu_wsph_convective_diff_tavgs}
\end{figure}

As with the previous analysis, we assess the signal-to-noise and main-to-interaction ratios of each parameter.
In Fig.~\ref{fig:snr_ysu_wsph_convective_diff_tavgs}, the signal-to-noise ratios reveal that signals of the parameters $b$, $p$, and $z_0$ in wind speed predictions are expected to be observable over the observational noise regardless of the time-averaging window.
In Fig.~\ref{fig:mir_ysu_wsph_convective_diff_tavgs}, the main-to-interaction ratios of these select parameters reveal that parameters $p$ and $z_0$ are practically identifiable from observations of wind speed across all four time-averaging windows over much of the PBL.
Here, we note a significant increase in the main-to-interaction ratio of $z_0$, which indicates that the independent-effect contribution increases with respect to interaction-effect contributions to the variance in model output (wind speed) with longer temporal averaging.
Although not shown here, this corresponds to an increase in the main-effect sensitivity of $z_0$, which is the independent contribution of the parameter to the variance of the model output.
In other words, the informativeness of observations for $z_0$ is expected to increase with longer temporal averaging.

\begin{figure}
    \centering
        \includegraphics[width=1.0\textwidth]{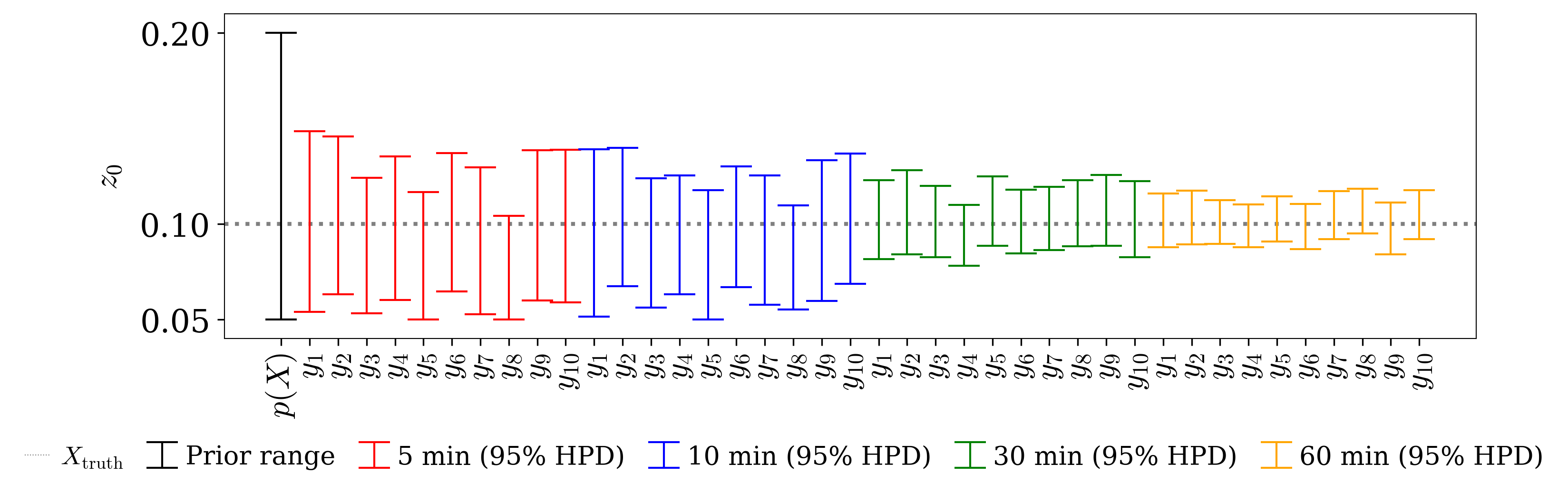}
    \caption{
    95\% highest posterior density (HPD) credible intervals of the roughness length $z_0$ inferred from wind-speed observations in the convective regime for four time-averaging windows: (a) 5 min (red), (b) 10 min (blue), (c) 30 min (green), and (d) 60 min (orange). 
    The prior range of $z_0$ is shown by the black error bar, and the truth parameter value is indicated by the dotted line. 
    Parameter inference is repeated for ten independent observational samples to quantify sample variability.
    }
    \label{fig:posterior_ysu_convective_wsph_z0_95hpd_diff_tavgs}
\end{figure}

This is corroborated by the marginal posterior distributions of $z_0$ learned from observations of convective-regime wind speed over the different time-averaging windows.
Figure~\ref{fig:posterior_ysu_convective_wsph_z0_95hpd_diff_tavgs} illustrates the $95\%$ highest posterior density (HPD) intervals of $z_0$ learned from ten independent samples of wind speed observations in order to test robustness to sample variability.
First, we observe a clear trend of the HPD intervals decreasing with longer temporal averaging.
However, this decrease is not linear with the averaging timescale.
In fact, although the uncertainty--quantified by the width of the 95\% HPD intervals--is comparable between 5- (red) and 10-minute (blue) averaging, and similarly between 30- (green) and 60-minute (orange) averaging, it decreases by 46\% on average when increasing the averaging window from 10 to 30 minutes.
This reduction suggests that the integral timescale of the flow lies between these two averaging intervals.
Reducing uncertainty in estimates of the roughness length, $z_0$, is essential, as $z_0$ cannot be measured directly and is known to exhibit estimate discrepancies of up to two orders of magnitude \cite{wiernga1993representative}.
We note that uncertainty in inferred $z_0$, and more broadly in other parameters, can be practically reduced by accounting for the temporal averaging needed to isolate scale-dependent phenomena in the flow.

\subsubsection{Influence of observed quantity of interest}
\label{sec:influenceofobservedquantityofinterest}

Next, we investigate the influence of the observed quantity of interest to the inferred parameter uncertainty.
The observed quantity is another user choice that may significantly affect how well the observations constrain parameters in a parameterization \cite{lamb2026perspectives}.
Once constrained at the process level, a parameterization must operate consistently when coupled with other parameterizations in an ESM and make accurate predictions for various applications \cite{stensrud2007parameterization}.
As such, we explore how the observed quantity influences the inferred parameter uncertainty.
Here, we focus on wind speed, wind direction, and potential temperature, but the method is generalizable to any modeled quantity.
As before, we isolate the effect of the quantity by fixing the other three factors to 30-minute averaged convective-regime statistics in five spatial locations uniformly up to the PBL height.

\begin{figure}
    \centering
        \includegraphics[width=0.85\textwidth]{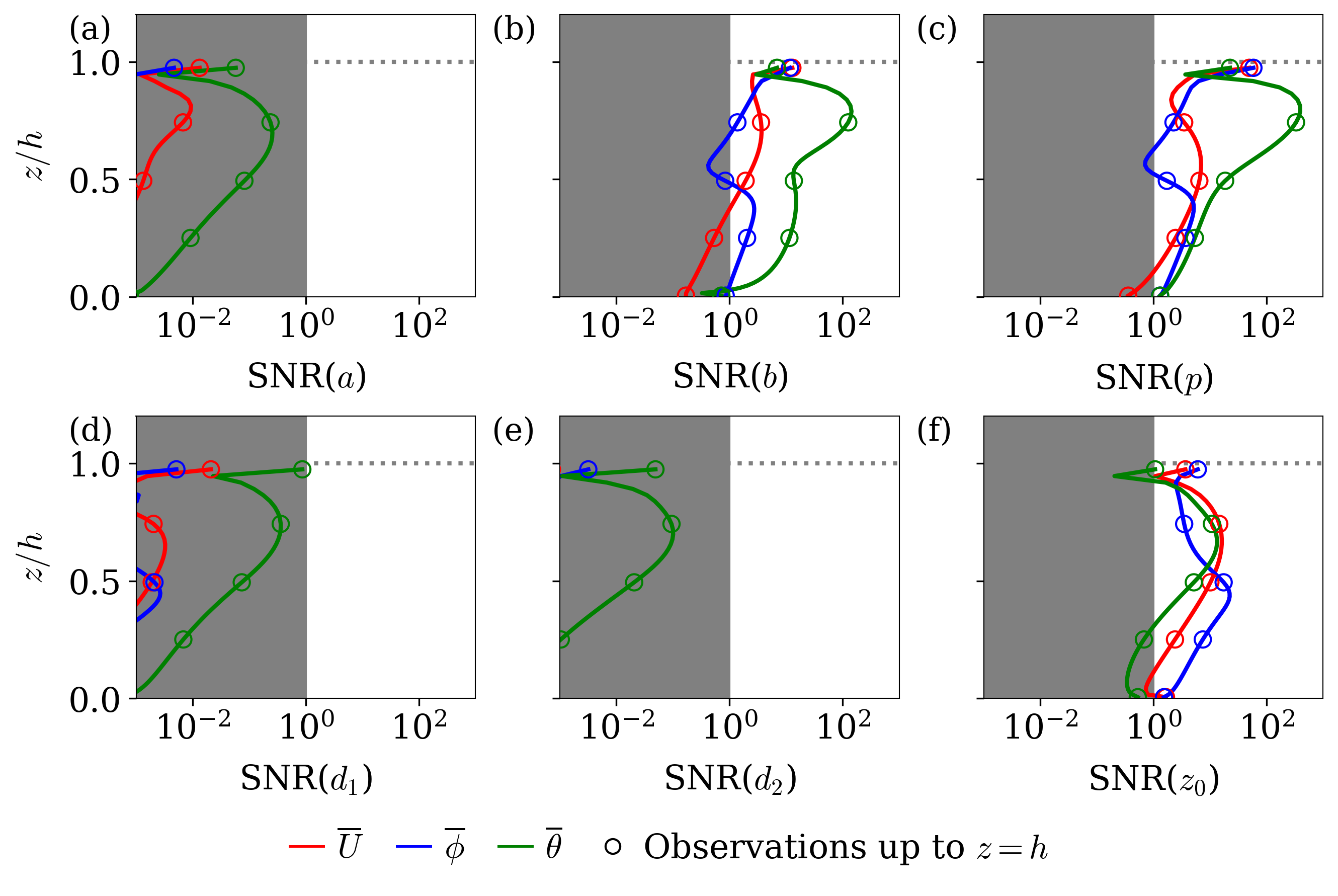}
    \caption{
    Signal-to-noise (SNR) ratios computed for the coupled YSU PBL-revised MM5 scheme parameters based on 30-minute-averaged quantities of interest in the convective regime: wind speed (red), wind direction (blue), and potential temperature (green).
    Circle markers correspond to the observed vertical locations up to PBL height $h$
    }
    \label{fig:snr_ysu_convective_30min_diff_qois}
\end{figure}

The signal-to-noise ratios of the coupled YSU PBL-revised MM5 scheme parameters from observations of wind speed (red), wind direction (blue), and potential temperature (green) are shown in Fig.~\ref{fig:snr_ysu_convective_30min_diff_qois}.
For observations of wind speed and wind direction, we observe that signals of parameters $a$, $d_1$, and $d_2$ are not expected to be visible in the observational noise.
On the other hand, for observations of potential temperature, we see that $d_1$ is potentially visible as it falls on the unity line at the PBL height.
With more certainty, we observe that the signals of parameters $b$, $p$, and $z_0$ are expected to be visible, regardless of observed quantity.
We also see how the signal-to-noise ratio may change as a function of the vertical location.
For example, the signal-to-noise ratios of $b$ for wind speed and $z_0$ for potential temperature, respectively, fall below the unity line near the surface $z=0$.
The identification of such regions in the observation space prior to the inference process is expected to be critical to the success of parameter calibration.

\begin{figure}
    \centering
        \includegraphics[width=0.85\textwidth]{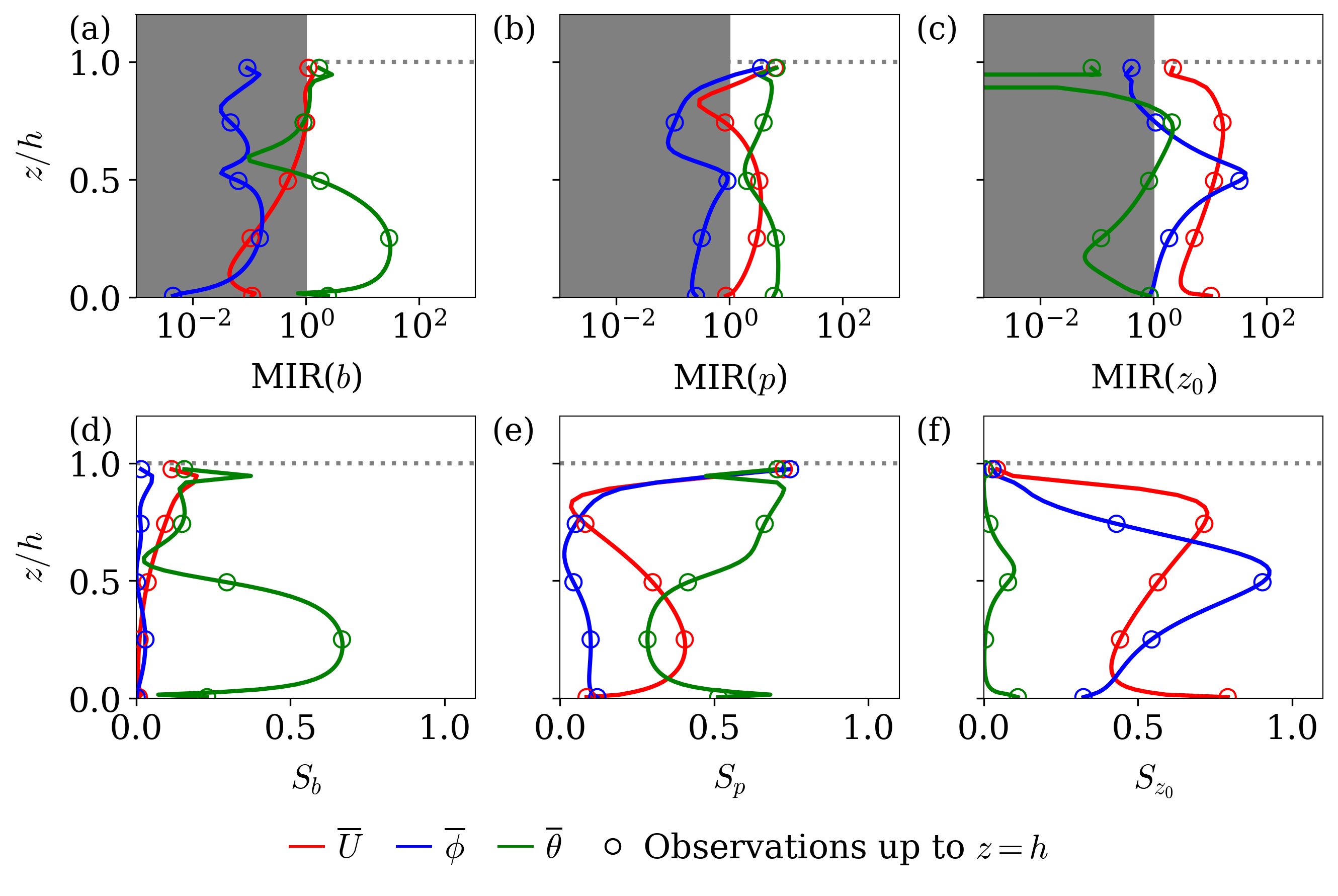}
    \caption{
    Main-to-interaction (MIR) ratio and main-effect sensitivity computed for the coupled YSU PBL-revised MM5 scheme parameters retained after filtering based on the corresponding signal-to-noise ratios, based on 30-minute-averaged quantities of interest in the convective regime: wind speed (red), wind direction (blue), and potential temperature (green).
    Circle markers correspond to the observed vertical locations up to PBL height $h$
    }
    \label{fig:mir_S1_ysu_convective_30min_diff_qois}
\end{figure}

The corresponding main-to-interaction ratios shown in Fig.~\ref{fig:mir_S1_ysu_convective_30min_diff_qois}a-c show that the criteria for practical identifiability of parameters $b$, $p$, and $z_0$ are both quantity- and location-dependent.
For example, we observe that parameter $b$ is expected to be dominated by interaction effects for wind direction observations across the PBL and therefore, practically non-identifiable.
On the other hand, the potential for uncertainty reduction is relevant to the upper bound of informativeness of observations, which we diagnose using the main-effect sensitivity of parameters.
The main-effect sensitivity of the select parameters for the different quantities of interest (Fig.~\ref{fig:mir_S1_ysu_convective_30min_diff_qois}d-f) visualize regions in the observation space where the parameters are expected to contribute most to the variance of the model output.
For example, parameter $b$ independently contributes most to the variance in the modeled potential temperature in the lower half of the PBL.
On the other hand, parameter $p$ contributes most to the variance in the modeled wind speed and wind direction at the PBL height.
Interestingly, roughness length $z_0$ contributes most strongly to the modeled wind speed near the surface and in the upper mixed layer, whereas its largest contribution to the modeled wind direction occurs in the middle of the mixed layer.
These are the regions in the observation space that are expected to result in the largest parameter posterior uncertainty reduction.

\begin{figure}
    \centering
        \includegraphics[width=1.0\textwidth]{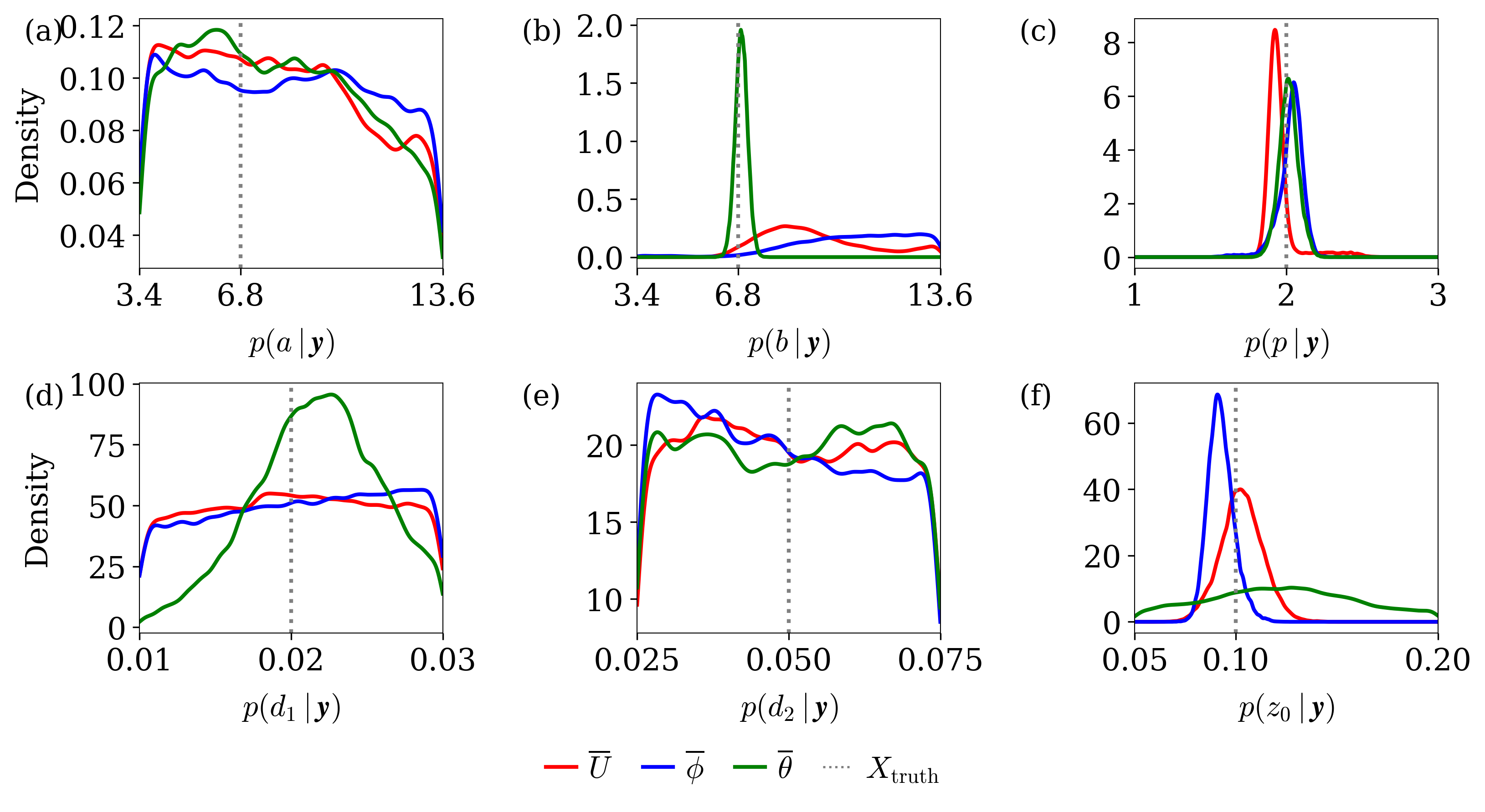}
    \caption{Marginal posterior distributions of the coupled YSU PBL-revised MM5 scheme parameters inferred from 30-minute-averaged quantities of interest in the convective regime: wind speed (red), wind direction (blue), and potential temperature (green). The truth parameter value is indicated by the dotted line.}
    \label{fig:posterior_ysu_convective_30min_diff_qois}
\end{figure}

Marginal posterior distributions of the coupled YSU PBL-revised MM5 scheme parameters are inferred from 30-minute-averaged convective-regime observations of the three different quantities of interest (Fig.~\ref{fig:posterior_ysu_convective_30min_diff_qois}).
Inferred posterior distributions highlight clear advantages of preferentially observing certain quantities to constrain certain parameters--for example, wind speed and wind direction for inferring the profile-shape exponent $p$, and potential temperature for inferring the countergradient term constant $b$.
Main-effect sensitivity indices are obtained through forward uncertainty propagation, whereas posterior distributions are inferred through Bayesian inference; therefore, there is no direct one-to-one correspondence between them.
However, we observe that the trends in main-effect sensitivity by the different quantities of interest are closely correlated with the trends observed in corresponding posterior distributions, suggesting that the former serve as a strong proxy for the latter.
Because these advantages may be assessed using the diagnostic measures of the Sobol' sensitivity indices before the inversion process, our framework also guides which quantities should be observed so that otherwise weakly constrained parameters become practically identifiable for calibration.

\subsubsection{Influence of spatial location of observations}
\label{sec:influenceofspatiallocationofobservations}

The spatial placement of observations has useful implications for measurement acquisition campaigns and data assimilation strategies \cite{bechmann2020most, chen2022optimal}.
From the investigation on the influence of observed quantity, we observed the importance of the spatial locations of observations to the inference process.
For example, the marginal posterior distribution of $z_0$ is smallest for wind direction because it is observed at its peak sensitivity location (see Fig.~\ref{fig:mir_S1_ysu_convective_30min_diff_qois}).
The same is true for parameter $d_1$, for which we observe a slight narrowing of the posterior distribution because we are observing potential temperature at the PBL height where the signal-to-noise ratio is approximately unity ($\mathrm{SNR}(i) \approx 1$) (see Fig.~\ref{fig:snr_ysu_convective_30min_diff_qois}).
Therefore, we investigate the influence of spatial location of observations to the inferred parameter uncertainty.
We return to the analogy of meteorological masts as a practical source of observations within the PBL.
Although meteorological masts can reach heights of up to approximately $500~\mathrm{m}$, many are less than $100~\mathrm{m}$ tall \cite{ramon_tall_2020}.
To compare against the previously specified scenario in which observations extend to the PBL height which is approximately $1000~\mathrm{m}$ \cite{shin2016evaluation}, we consider a more realistic observation scenario with five measurements distributed uniformly up to $250~\mathrm{m}$ \cite{pena2020evaluating}.
Here, the influence of spatial locations across different quantities of interest is explored by fixing the two other factors to 30-minute-averaged quantities observed in the convective regime.

\begin{figure}
    \centering
        \includegraphics[width=0.87\textwidth]{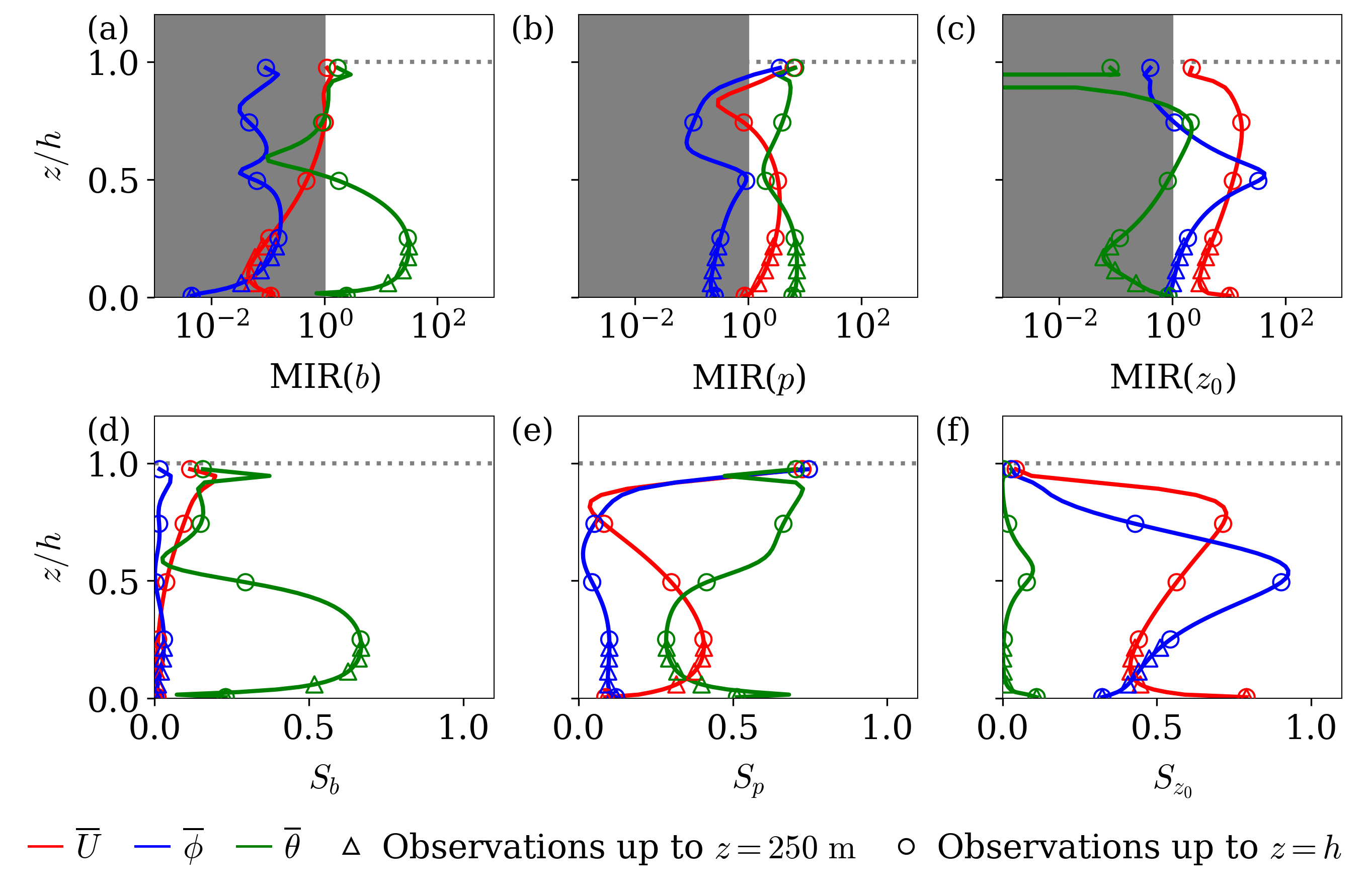}
    \caption{
    Main-to-interaction (MIR) ratio and main-effect sensitivity computed for the coupled YSU PBL-revised MM5 scheme parameters retained after filtering based on the corresponding signal-to-noise ratios, based on 30-minute-averaged quantities of interest in the convective regime: wind speed (red), wind direction (blue), and potential temperature (green)
    Triangle markers correspond to observed vertical locations up to $250~\mathrm{m}$.
    Circle markers correspond to the observed vertical locations up to PBL height $h$.
    }
    \label{fig:mir_S1_ysu_convective_30min_diff_qois_locs}
\end{figure}

The main-to-interaction ratio and main-effect sensitivity of parameters for the different modeled quantities that were shown in Fig.~\ref{fig:mir_S1_ysu_convective_30min_diff_qois} are updated with the observed locations up to $250~\mathrm{m}$ (triangle markers) in Fig.~\ref{fig:mir_S1_ysu_convective_30min_diff_qois_locs}.
For observations up to $250~\mathrm{m}$, we now expect countergradient term constant $b$ to only be practically identifiable from observations of potential temperature.
On the other hand, we expect profile-shape exponent $p$ to be practically identifiable from observations of wind speed and potential temperature.
The main-to-interaction ratios show that the interaction contributions of parameters for the other quantities are higher than the independent contributions.

\begin{figure}
    \centering
        \includegraphics[width=1.0\textwidth]{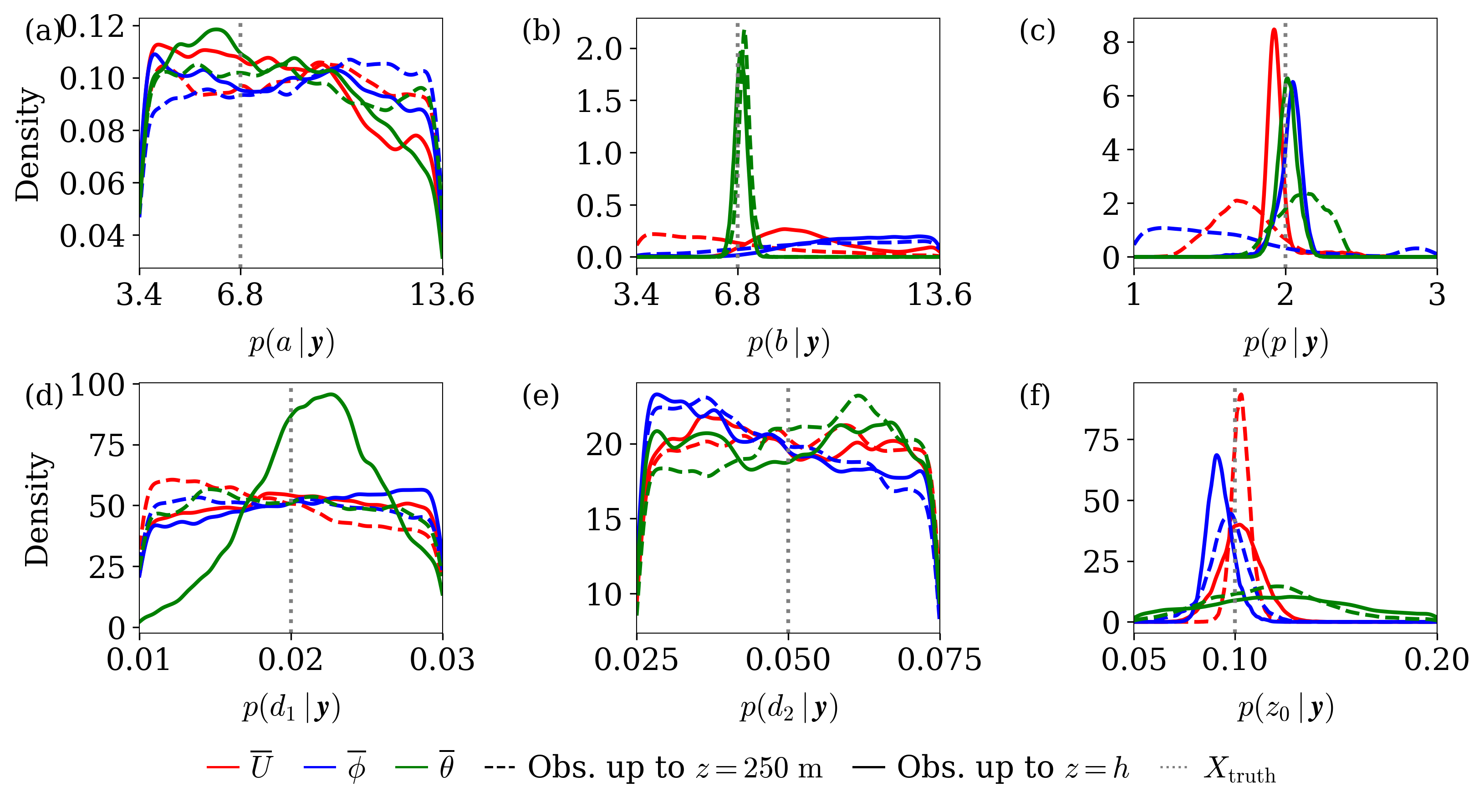}
    \caption{Marginal posterior distributions of the coupled YSU PBL-revised MM5 scheme parameters inferred from 30-minute-averaged quantities of interest in the convective regime: wind speed (red), wind direction (blue), and potential temperature (green). The results are based on observations extending to $250~\mathrm{m}$ (dashed) or observations extending to $h$ (solid). The truth parameter value is indicated by the dotted line.}
    \label{fig:posterior_ysu_convective_30min_diff_qois_heights}
\end{figure}

In Fig.~\ref{fig:posterior_ysu_convective_30min_diff_qois_heights}, we show marginal posterior distributions of parameters inferred from wind speed, wind direction, and potential temperature for two spatial-location scenarios: five vertical locations distributed up to $250~\mathrm{m}$ (dashed) and five vertical locations distributed up to the PBL height $h$ (solid).
In reference to the observation-specific identifiability of parameter $d_1$ mentioned previously, we see that $d_1$ is not constrained by observations are up to $250~\mathrm{m}$.
Furthermore, when observations are limited to heights below 250 m, the true value of parameter $p$ lies outside approximately 90\% highest posterior density region inferred from wind direction, and the posterior mode shows substantial bias (see Fig.~\ref{fig:snr_ysu_convective_30min_diff_qois}).
We attribute this miscalibration that occurs despite the signal-to-noise ratio exceeding unity to the main-to-interaction ratio where the interaction effects are larger than the main-effect contributions in the locations observed (see Fig.~\ref{fig:mir_S1_ysu_convective_30min_diff_qois_locs}b).
This is an example where although the parameter signal is visible over observational noise, it is a region of the observation space where the interaction effects of parameters are dominant.
This is in contrast to the scenario in which wind direction is observed up to the PBL height, where we show that the calibration of $p$ is significantly improved with respect to the truth parameter value.
A physical insight is that the PBL height is an important location to sample for key eddy-viscosity-controlling parameters \cite{nielsen2010evaluation,baki2022determining,shin2026accelerated}.
For parameter $b$, we observe that the lower $250~\mathrm{m}$ is the region where its main-effect sensitivity is highest for modeled potential temperature.
This is corroborated by the negligible change in posterior distribution between the two spatial-location scenarios.
Lastly, we observe that the credible intervals of $z_0$ are significantly reduced for observations up to $250~\mathrm{m}$, although this is not explained by the main-effect sensitivity.

\FloatBarrier

\section{Conclusions}
\label{sec:conclusions}

The current study demonstrates an efficient framework, aided by machine-learning emulators, that connects the forward problem to the inverse problem of UQ using diagnostic measures of sensitivity indices for parameter calibration in ESM parameterizations.
We introduced that the framework informs a non-iterative selection of observations considering multiple factors in constructing informative observations for a chaotic dynamical system.
In doing so, we showed that modular, lightweight ML emulators make it practical to apply standard statistical analyses to gain physical insight into ESM parameterizations and use that insight to guide their calibration.

Global sensitivity analysis of parameters provides a quantitative interpretation of how the model represents planetary boundary-layer physics, since the calibrated parameters correspond to physically meaningful processes governing turbulent momentum and heat transport.
We revealed how the parameter sensitivities depend on the output quantity of interest, time-averaging window, atmospheric stability regime, and spatial locations.
For wind speed, parameters controlling turbulent production and dissipation--such as the profile-shape exponent $p$ in the YSU PBL scheme--primarily influence the region near the PBL height. 
In contrast, the roughness length $z_0$, which enters the PBL scheme through surface fluxes via the surface-layer parameterization, mainly affects the surface layer and, depending on the stability regime, the flow aloft.
For potential temperature, the effect of $z_0$ decreases as parameters relevant to buoyancy such as the countergradient term constant $b$ increase in contribution.
Parameter sensitivities are more strongly affected by temporal averaging in the convective regime than in the stable regime, likely due to the larger turbulence scales of convective flow.
Recognizing how these factors that are relevant to the chaotic dynamical system interact can inform how model parameters should be calibrated using observations.
Additionally, the interaction-effect sensitivities of parameters create an opportunity for a more systematic exploration of parameter interactions arising from coupled physics parameterizations in ESMs.

In addition to revealing parameter sensitivities of the model, we introduced a framework that leverages information from the sensitivity analysis to systematically and non-iteratively guiding the selection of observations for Bayesian model calibration of ESM parameterizations.
We assert that the parameter sensitivity projected onto the observation space is the primary indicator for the informativeness an observation for parameter calibration.
This motivates our framework which uses nondimensional diagnostic measures of sensitivity indices to identify regions in the observation space with relatively low observational noise and weak parameter interaction effects, prior to the inference process.
This possibility of a non-iterative selection of informative observations provides an efficient pathway to improving observational efficiency in Bayesian learning.
By performing emulator-aided Bayesian inversion with synthetic observations for which the ground-truth parameter values are known, we demonstrated how the uncertainty across the coupled YSU PBL-revised MM5 scheme parameters can be systematically reduced by leveraging information on parameter sensitivities to the observed quantity of interest, atmospheric stability during observation, observation averaging time length, and the spatial location of the observations.
In doing so, we corroborated our assertion that observations are informative to parameters in regions of the observation space where the model is sensitive to those target parameters.

These findings are motivated by the core idea that the informativeness of an observation for model calibration is a useful measure of its value to the model.
Therefore, understanding the model is the best way to identify the most informative observations.
In this sense, the forward and inverse statistical paradigms are best viewed as complementary in model calibration, where insights from the forward approach can improve the observational efficiency of Bayesian model inversion.
We expect this perspective may be useful further parameterization development and efficient observation use for Earth system models.

Future work should consider the application of the observation selection framework in non-idealized simulations initialized and forced with reanalysis data over realistic terrain and surface conditions.
The framework should be evaluated in settings of surface heterogeneity, where the observation space expands from vertical locations to include horizontal locations to address questions on land-atmosphere interactions over surface heterogeneity in ESMs.
Furthermore, efforts to expand on the framework with an explicit treatment of the model-form or structural error are needed.
This will allow learning and development of parameterizations from a wider range of observations, such as field observations, turbulence-resolving simulations, etc.
Finally, the framework should be used to investigate the influence of parameters across the coupled physics parameterizations, including microphysics and radiation schemes, to elucidate phenomena arising from the complex, coupled turbulent dynamics of the atmosphere.

%
%

\section*{Open Research Section}

All computer code used in this paper is open source.
The code for the Weather Research and Forecasting (WRF) Model v4.6.1 is available at: (https://github.com/Howland-Lab/WRF-4.6.1).
SALib, the open-source Python library for sensitivity analysis, is available at: (https://github.com/SALib/SALib.git).
The data used to generate the figures and the code used to run the UQ analyses are available at: (https://doi.org/10.5281/zenodo.19667387).

\section*{Conflict of Interest declaration}
The authors declare there are no conflicts of interest for this manuscript.

\acknowledgments
E. Y. S and M. F. H acknowledge support of Office of Naval Research, Young Investigator Program (YIP), grant no. N000142512045. 
E.Y.S. acknowledges support from the MIT MathWorks fellowship. 
Simulations were performed on the Stampede3 super-computers under the NSF EXPLORE project EES240114, NSF ACCELERATE project EVE240008, and NSF MAXIMIZE project ATM170028.

\FloatBarrier

\appendix
\renewcommand{\thesection}{Appendix \arabic{section}}

\section{Emulator assessment}
\label{sec:appendix1}

\begin{figure}
    \centering
        \includegraphics[width=0.85\textwidth]{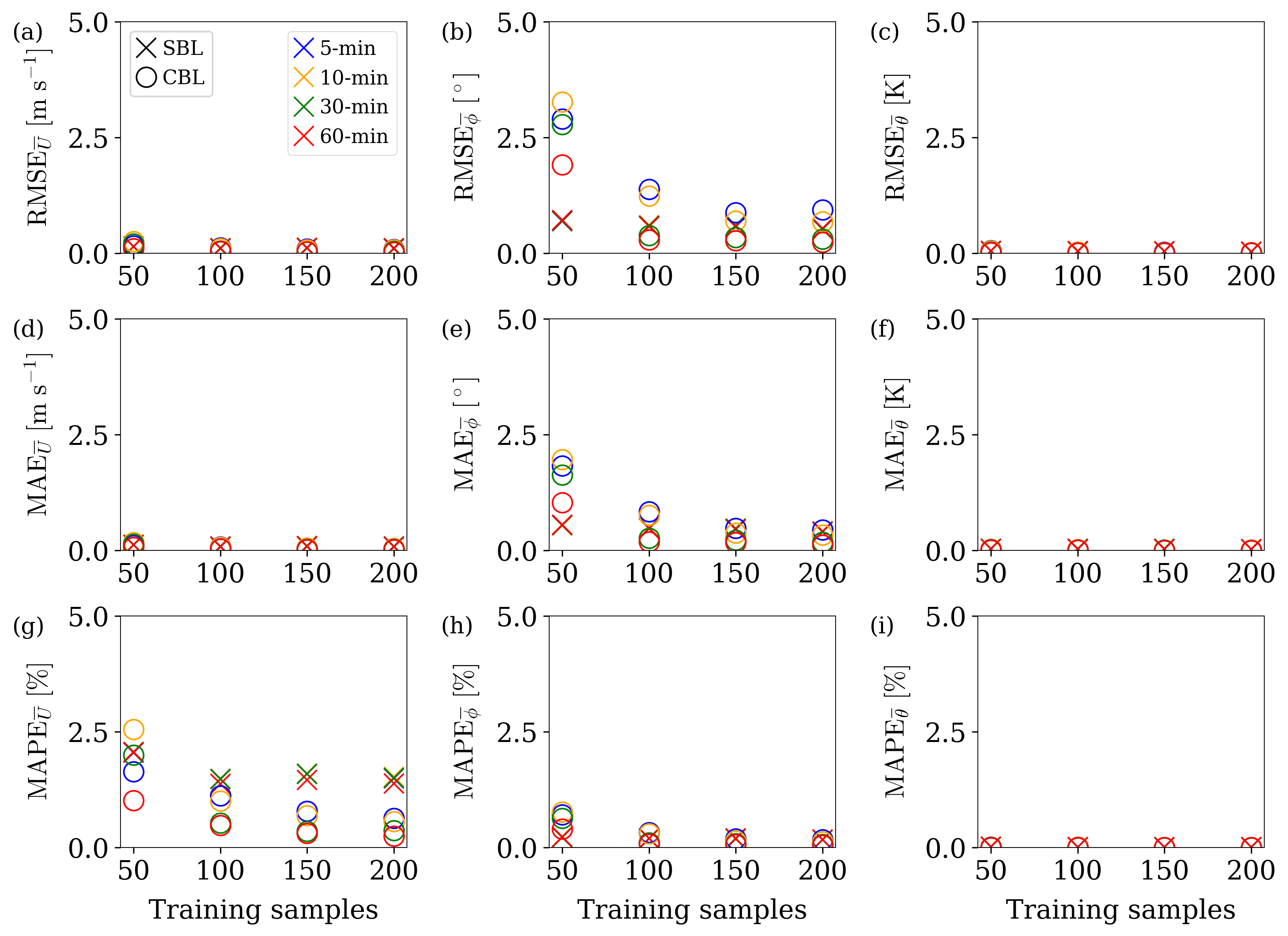}
    \caption{
    Root-mean-square error (RMSE), mean absolute error (MAE), and mean absolute percentage error (MAPE) of individual GP emulators that map YSU PBL scheme parameters to model outputs averaged over different temporal windows for (a,d,g) wind speed, (b,e,h) wind direction, and (c,f,i) potential temperature in the PBL
    }    
    \label{fig:ysu_gp_assessment_results}
\end{figure}

We evaluate the performance of the Gaussian process (GP) emulators using standard error metrics, including root-mean-square error (RMSE), mean absolute error (MAE), and mean absolute percentage error (MAPE), on an 80-20 train-test split (Fig.~\ref{fig:ysu_gp_assessment_results}). 
These metrics are computed across output quantities of interest, time-averaging windows, and atmospheric stability regimes, with model outputs averaged over vertical levels within the PBL.
To assess the required number of training samples of true WRF model evaluations, we vary the number of WRF simulations used to train the emulators. 
We find that the error metrics converge across all quantities of interest after approximately 100 training samples, which indicates that additional model evaluations yield diminishing improvements in accuracy.
Overall, the emulators achieve low RMSE, MAE, and MAPE across wind speed, wind direction, and potential temperature.
We observe that the maximum MAPE is below 2.5\% for wind speed and well below 1\% for both wind direction and potential temperature. 
These results indicate that the emulators provide an accurate surrogate of the true model response over the parameter space.

\begin{figure}
    \centering
        \includegraphics[width=0.85\textwidth]{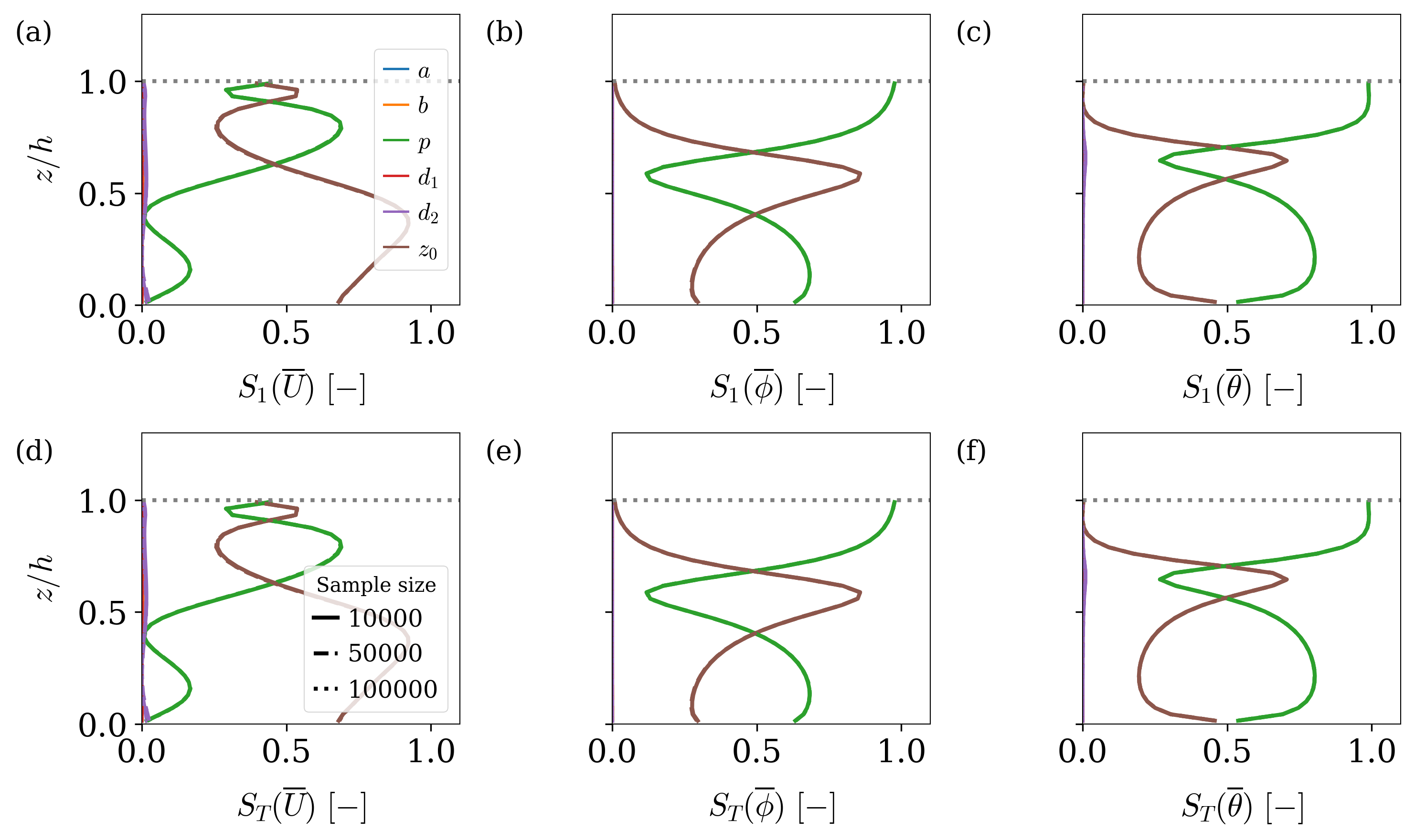}
    \caption{
    Sobol' indices versus MCMC sample size for the emulator of the YSU PBL scheme for 30-minute-averaged statistics in the stable boundary layer. 
    Panels (a–c) report main-effect indices for (a) wind speed, (b) wind direction, and (c) potential temperature. 
    Panels (d–f) report total-effect indices for the same outputs. 
    Increasing $S_1$ ($S_T$) indicates a stronger main (total) contribution of the corresponding parameter; trends with sample size show estimator convergence.
    }
    \label{fig:ysu_stable_sobol_indices_sensitivity_to_sample_size}
\end{figure}

\begin{figure}
    \centering
        \includegraphics[width=0.85\textwidth]{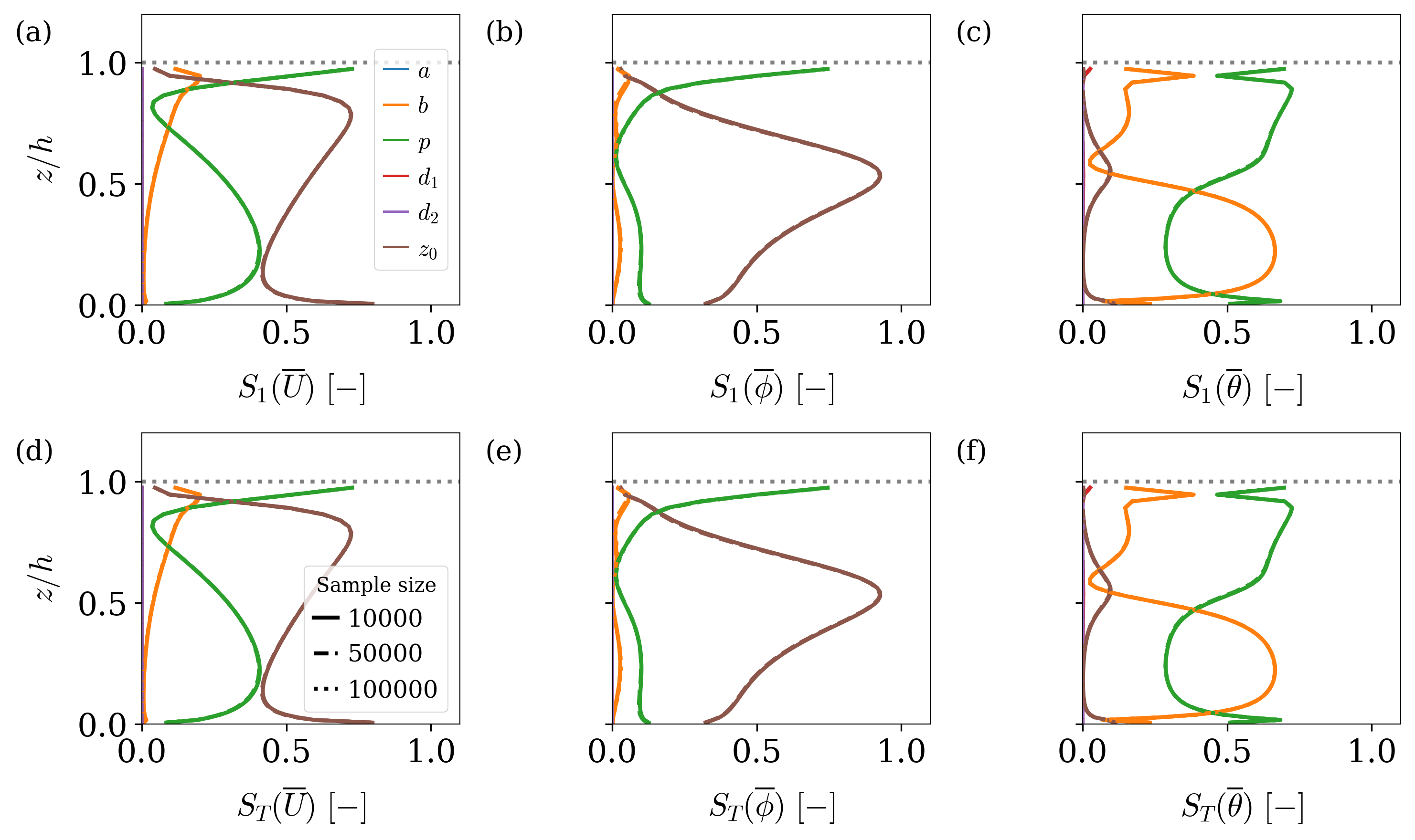}
    \caption{
    Sobol' indices versus MCMC sample size for the emulator of the YSU PBL scheme for 30-minute-averaged statistics in the convective boundary layer. 
    Panels (a–c) report main-effect indices for (a) wind speed, (b) wind direction, and (c) potential temperature. 
    Panels (d–f) report total-effect indices for the same outputs. 
    Increasing $S_1$ ($S_T$) indicates a stronger main (total) contribution of the corresponding parameter; trends with sample size show estimator convergence.
    }
    \label{fig:ysu_convective_sobol_indices_sensitivity_to_sample_size}
\end{figure}

We also make sure that the GP emulators are suitable for global sensitivity analysis by assessing the convergence of Sobol’ indices with respect to the number of Markov chain Monte Carlo (MCMC) samples.
Figures~\ref{fig:ysu_stable_sobol_indices_sensitivity_to_sample_size} and \ref{fig:ysu_convective_sobol_indices_sensitivity_to_sample_size} show that both main-effect and total-effect sensitivity indices are invariant to the sufficiently large sample sizes for stable and convective boundary layer conditions, respectively. 
This indicates that the emulator-based estimates of Sobol’ indices are reliable and that the sample size ($50{,}000$) used in the main results is sufficient to quantify parameter sensitivities in both stability regimes.

\FloatBarrier

\section{Supplementary sensitivity analysis results}
\label{sec:appendix2}

We also conduct global sensitivity analysis on two additional PBL schemes: Mellor–Yamada–Nakanishi–Niino Level 2.5 (MYNN 2.5) \cite{nakanishi2009development} and Shin \& Hong (SH) \cite{shin2015representation}.
Both PBL schemes are coupled with the revised version of the fifth-generation Pennsylvania State University-National Center for Atmospheric Research Mesoscale Model (MM5) SFCLAY scheme \cite{jimenez2012revised}.
The parameterizations and their targeted parameters are listed in Table~\ref{tab:sl-pbl-parameters_apdx}.
The uniform distributions used are define the ranges over which parameters are randomly sampled for UQ and are informed by previous studies \cite{yang2017sensitivity,baki2022determining}.

\begin{table}[t]
\centering
\caption{Target parameters in the planetary boundary layer (PBL) and surface-layer (SFCLAY) schemes}
\label{tab:sl-pbl-parameters_apdx}
\begin{tabular}{lllll}
\hline\noalign{\smallskip}
Parameterization type & Scheme & Parameter & Standard value & Range \\
\hline\noalign{\smallskip}
\multirow{5}{*}{Planetary boundary layer}
& \multirow{5}{*}{MYNN 2.5}
& $g_1$     & 0.235 & $\mathcal{U}(0.1768, 0.2395)$ \\
& & $b_1$     & 24    & $\mathcal{U}(12, 36)$ \\
& & $c_3$     & 0.34  & $\mathcal{U}(0.33, 0.50)$ \\
& & $c_5$     & 0.2   & $\mathcal{U}(0.1, 0.3)$ \\
& & $cns$     & 2.7   & $\mathcal{U}(1.35, 4.05)$ \\
& & $\beta$   & 0.2   & $\mathcal{U}(0.1, 0.3)$ \\
& \multirow{5}{*}{Shin \& Hong}
& $a$       & 6.8  & $\mathcal{U}(3.4,13.6)$ \\
& & $b$     & 6.8  & $\mathcal{U}(3.4,13.6)$ \\
& & $p$     & 2    & $\mathcal{U}(1,3)$ \\
& & $d_{1}$ & 0.02 & $\mathcal{U}(0.01,0.03)$ \\
& & $d_{2}$ & 0.05 & $\mathcal{U}(0.025,0.075)$ \\
Surface layer & Revised MM5
& $z_{0}$ & 0.1 & $\mathcal{U}(0.05,0.20)$ \\
\noalign{\smallskip}\hline
\end{tabular}
\end{table}

\begin{figure}
    \centering
        \includegraphics[width=0.85\textwidth]{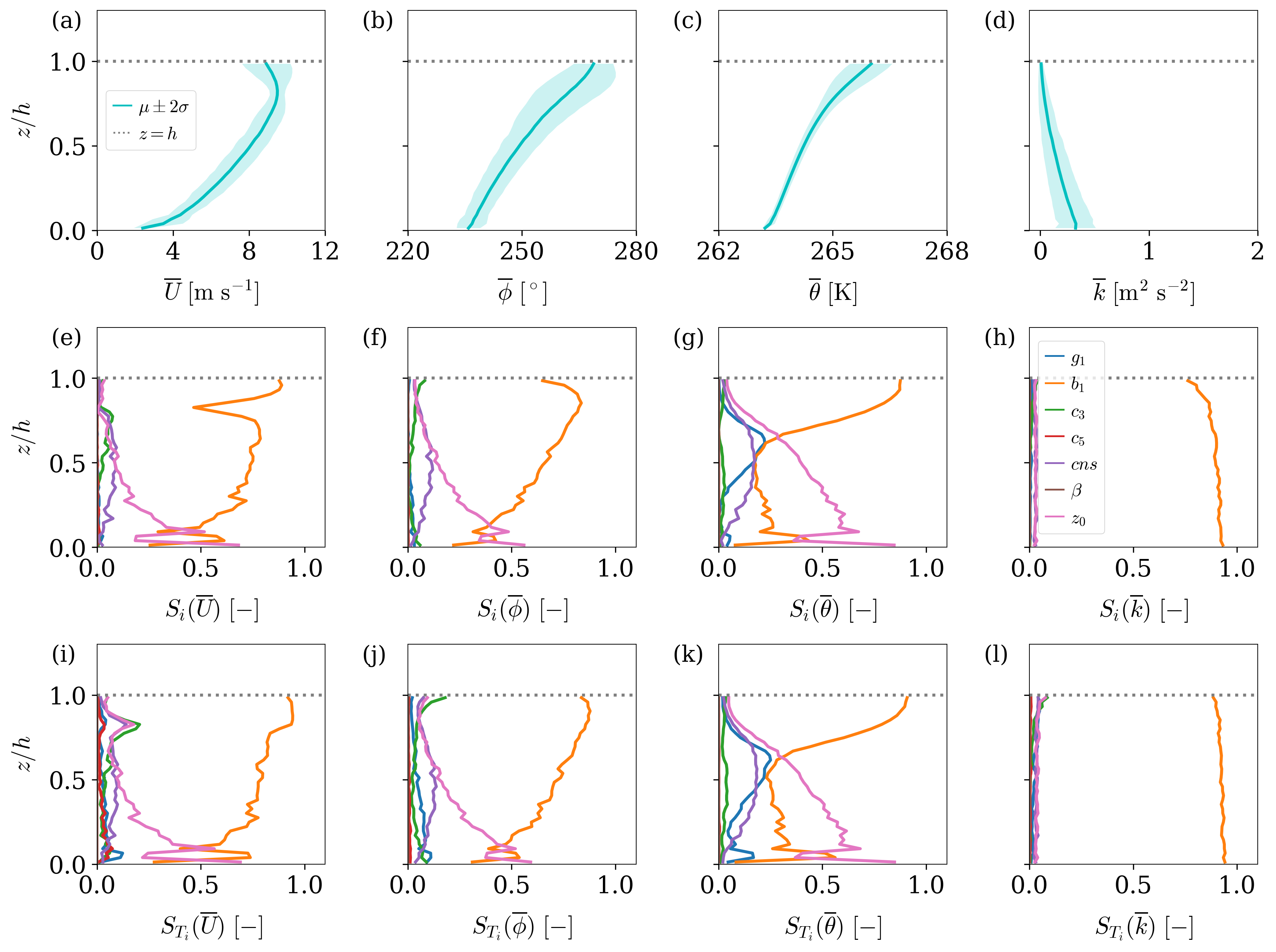}
    \caption{30-minute-averaged profiles of (a) wind speed, (b) wind direction, (c) potential temperature, and (d) turbulence kinetic energy in the stable boundary layer: (e–h) main-effect and (i–l) total-effect sensitivity indices of parameters in the MYNN Level 2.5 PBL scheme}
    \label{fig:sobol_indices_mynn_stable_30min_diff_qois}
\end{figure}

\begin{figure}
    \centering
        \includegraphics[width=0.85\textwidth]{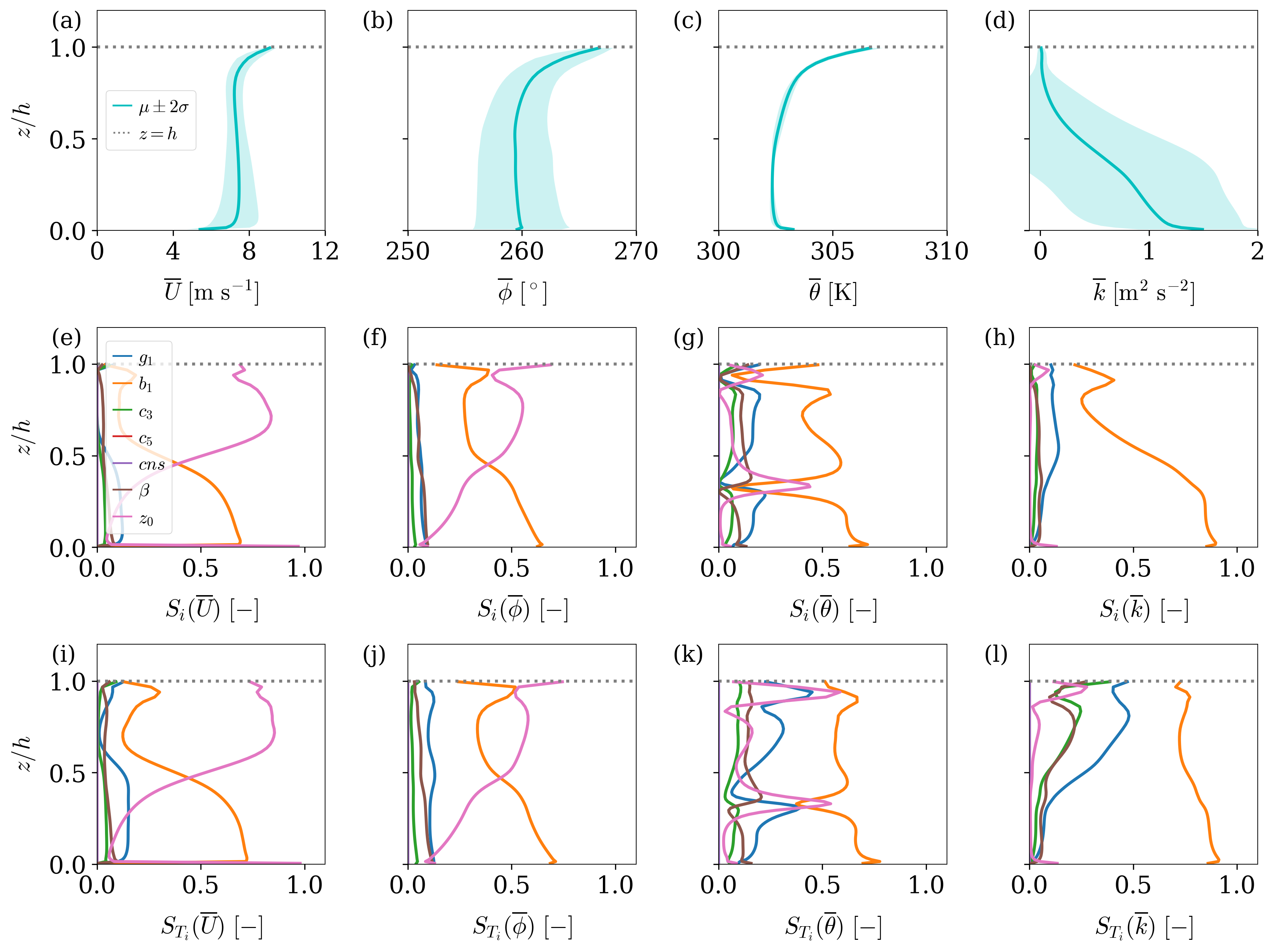}
    \caption{30-minute-averaged profiles of (a) wind speed, (b) wind direction, (c) potential temperature, and (d) turbulence kinetic energy in the convective boundary layer: (e–h) main-effect and (i–l) total-effect sensitivity indices of parameters in the MYNN Level 2.5 PBL scheme}
    \label{fig:sobol_indices_mynn_convective_30min_diff_qois}
\end{figure}

\begin{figure}
    \centering
        \includegraphics[width=0.85\textwidth]{figures/appendix/sobol_indices_sh_stable_30min_diff_qois.png}
    \caption{30-minute-averaged profiles of (a) wind speed, (b) wind direction, (c) potential temperature, and (d) turbulence kinetic energy in the stable boundary layer: (e–h) main-effect and (i–l) total-effect sensitivity indices of parameters in the Shin\& Hong PBL scheme}
    \label{fig:sobol_indices_sh_stable_30min_diff_qois}
\end{figure}

\begin{figure}
    \centering
        \includegraphics[width=0.85\textwidth]{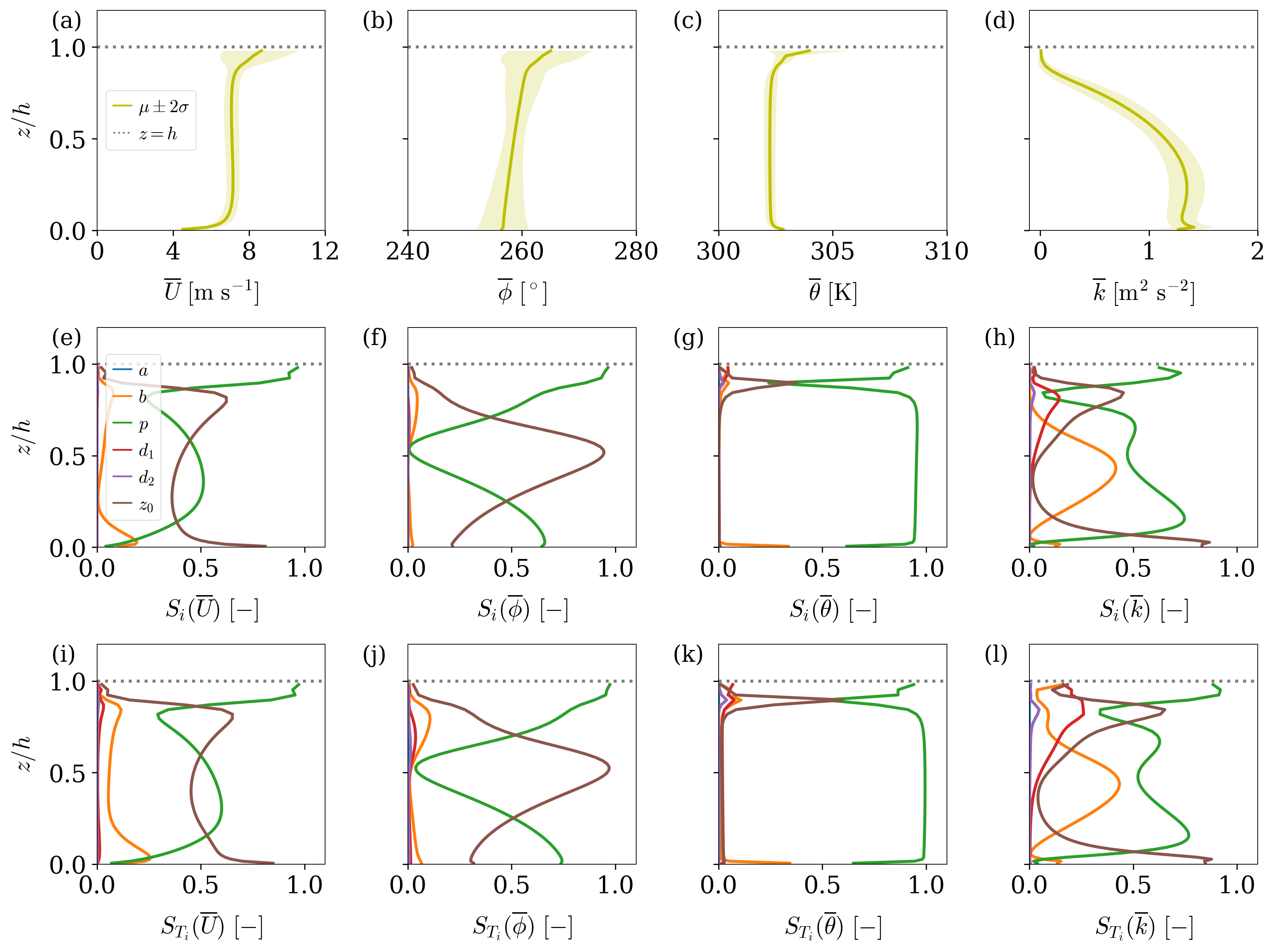}
    \caption{30-minute-averaged profiles of (a) wind speed, (b) wind direction, (c) potential temperature, and (d) turbulence kinetic energy in the convective boundary layer: (e–h) main-effect and (i–l) total-effect sensitivity indices of parameters in the Shin \& Hong PBL scheme}
    \label{fig:sobol_indices_sh_convective_30min_diff_qois}
\end{figure}

\textbf{Mellor–Yamada–Nakanishi–Niino Level 2.5 (MYNN 2.5)} PBL scheme is a 1.5-order, turbulence kinetic energy (TKE)-based local turbulence closure \cite{nakanish2001improvement}.
The level 2.5 variant predicts TKE with a prognostic equation but uses closure assumptions to predict other second-moment quantities diagnostically.
We target the parameter set $\boldsymbol{X}_{\mathrm{MYNN}}=\{g_1,\, b_1,\, c_3,\, c_5,\, cns,\, \beta,\, z_0\}$.
Several additional parameters exist, but we exclude them from the target parameter set based on the prior sensitivity analysis of \citeA{yang2017sensitivity}.
Parameter $g_1$ enters the formulations of several interconnected closure constants; to preserve parameter independence, those constants are not included in the parameter set.
Parameter $b_1$ is a coefficient in the formulation of the diagnostic dissipate rate.
Parameters $c_3$ and $c_5$ are coefficients in the parameterization of the pressure covariance.
Parameter $cns$ is a coefficient in the diagnostic turbulent length scale for shear production.
Parameter $\beta$ is an exponent constant in the diagnostic turbulent length scale for shear production in the convective regime.

\textbf{Shin \& Hong (SH)} PBL scheme is a grid-scale-aware variant of YSU PBL scheme that reduces parameterized local and non-local mixing as grid spacing approaches the gray zone, while retaining YSU’s first-order K-profile structure in stable conditions.
Therefore, under stable conditions, model outputs by the Shin \& Hong scheme are expected to be quantitatively similar to those by the YSU scheme.
We target the parameter set $\boldsymbol{X}_{\mathrm{SH}} = \{a, b, p, d_1, d_2, z_0\}$.
Parameter $a$ is a coefficient for the effect of the surface buoyancy flux on virtual potential temperature at the PBL height $h$.
Parameter $b$ is a coefficient for the effect of the countergradient mixing term.
Parameter $p$ is the exponent constant in the formulation of eddy diffusivity.
Parameter $d_1$ and $d_2$ are coefficients that formulate the thickness of the entrainment zone.
These parameters are formulated in the same way as in YSU scheme; however, because the mixing terms depend on grid spacing, the physical interpretation must account for this confounding effect as the horizontal grid spacing approaches the gray zone.
An advantage of this scheme over the YSU scheme is the availability of diagnostic turbulence variables, which are important higher-order moment outputs.

Figure~\ref{fig:sobol_indices_mynn_stable_30min_diff_qois} shows the parameter sensitivities in the MYNN PBL scheme for 30-minute-averaged statistics of a stable boundary layer: the first row shows the four output quantities of interest, followed by the corresponding main-effect and total-effect sensitivity indices.
Figure~\ref{fig:sobol_indices_mynn_convective_30min_diff_qois} shows corresponding parameter sensitivities in the MYNN PBL scheme for 30-minute-averaged statistics of a convective boundary layer.
Figure~\ref{fig:sobol_indices_sh_stable_30min_diff_qois} shows corresponding parameter sensitivities in the Shin \& Hong PBL scheme for 30-minute-averaged statistics of a stable boundary layer.
Figure~\ref{fig:sobol_indices_sh_convective_30min_diff_qois} shows corresponding parameter sensitivities in the Shin \& Hong PBL scheme for 30-minute-averaged statistics of a convective boundary layer.

\begin{figure}
    \centering
        \includegraphics[width=0.85\textwidth]{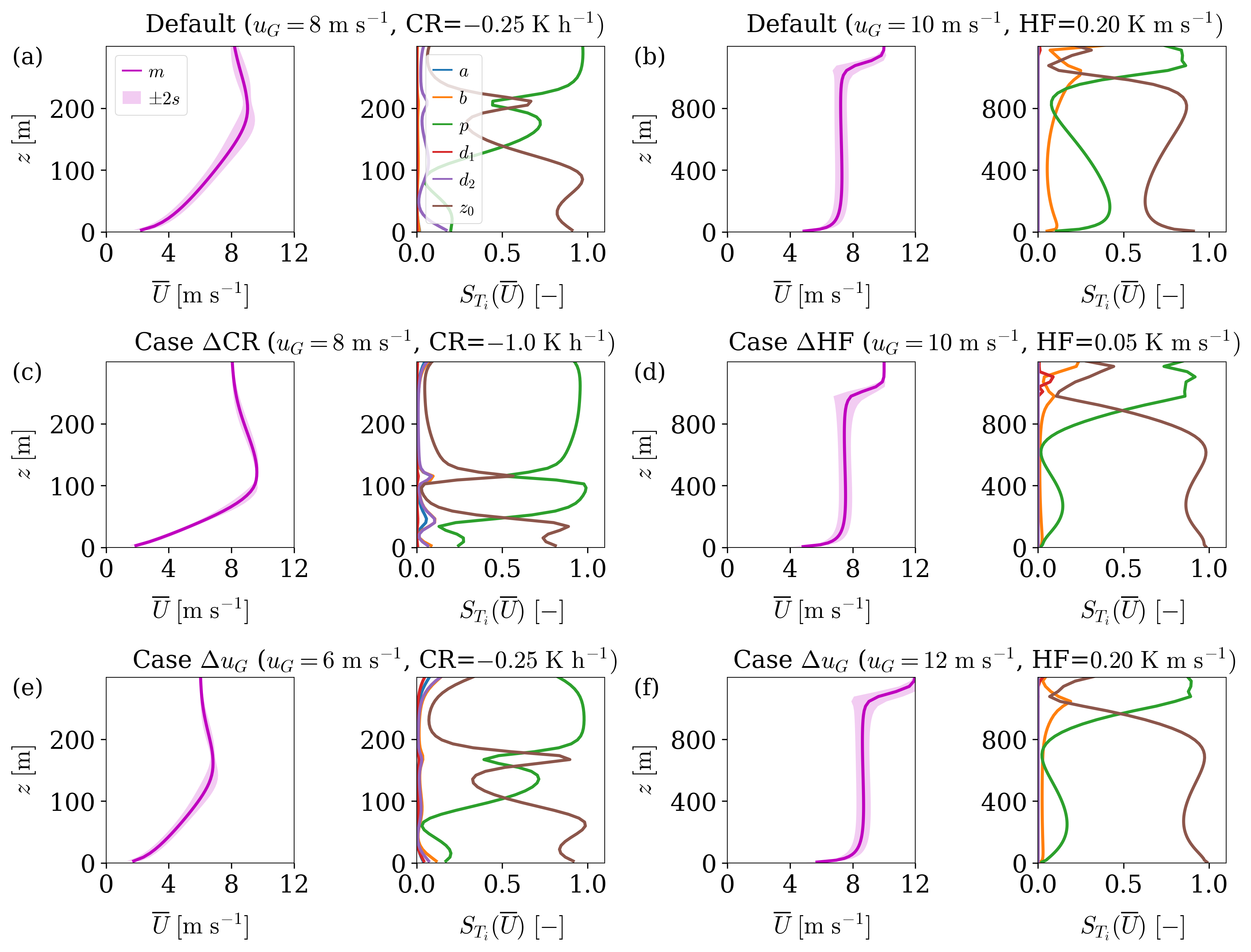}
    \caption{
    Wind-speed profiles and corresponding total-effect sensitivity indices for six cases:
    (a) $u_G=8~\mathrm{m~s^{-1}}$, cooling rate $=-0.25~\mathrm{K~h^{-1}}$ (stable);
    (b) $u_G=10~\mathrm{m~s^{-1}}$, surface heat flux $=0.20~\mathrm{K~m~s^{-1}}$ (convective);
    (c) $u_G=8~\mathrm{m~s^{-1}}$, cooling rate $=-1.0~\mathrm{K~h^{-1}}$ (stable);
    (d) $u_G=10~\mathrm{m~s^{-1}}$, surface heat flux $=0.05~\mathrm{K~m~s^{-1}}$ (convective);
    (e) $u_G=6~\mathrm{m~s^{-1}}$, cooling rate $=-0.25~\mathrm{K~h^{-1}}$ (stable);
    (f) $u_G=12~\mathrm{m~s^{-1}}$, surface heat flux $=0.20~\mathrm{K~m~s^{-1}}$ (convective).
    }
    \label{fig:sobol_indices_ysu_wsph_60min_diff_cases}
\end{figure}

We also conduct global sensitivity analysis of the YSU PBL scheme parameters under moderate changes in mesoscale and surface forcings within each stability regime (stable and convective) (Fig.~\ref{fig:sobol_indices_ysu_wsph_60min_diff_cases}). 
For each case, either the geostrophic wind magnitude $u_G$ or a surface forcing--represented by a cooling rate (CR) or heat flux (HF)--is systematically varied while maintaining a fixed stability regime.
Across these cases, the relative contributions of the parameters to the variance in modeled wind speed within the PBL remain largely consistent.

\FloatBarrier

\bibliography{agusample}

\end{document}